\documentclass[journal,twoside]{IEEEtran}

\ifCLASSINFOpdf
\else
\fi

\usepackage{amsmath, amsfonts, amssymb}
\usepackage{cite}
\usepackage{graphicx,graphics}
\usepackage{textcomp}
\usepackage{tikz}
\usetikzlibrary{arrows}
\usetikzlibrary{shapes}
\usepackage{tabulary}
\usepackage{array}
\usepackage{subcaption}
\usepackage{caption}
\usepackage{siunitx}
\usepackage{svg}
\usepackage{enumitem}
\usepackage{setspace}
\usepackage{MnSymbol}
\usepackage{mathtools}
\usepackage{scalerel}
\usepackage{algorithm,algorithmic}

\newcolumntype{L}[1]{>{\raggedright\let\newline\\\arraybackslash\hspace{0pt}}m{#1}}
\newcolumntype{C}[1]{>{\centering\let\newline\\\arraybackslash\hspace{0pt}}m{#1}}
\newcolumntype{R}[1]{>{\raggedleft\let\newline\\\arraybackslash\hspace{0pt}}m{#1}}

\DeclareMathOperator*{\argmin}{arg\,min}
\DeclareMathOperator*{\minimize}{minimize}

\newenvironment{smalleralign}[1][\small]
 {\par\nopagebreak\leavevmode\vspace*{-\baselineskip}%
  \skip0=\abovedisplayskip
  #1%
  \def\maketag@@@##1{\hbox{\m@th\normalfont\normalsize##1}}%
  \abovedisplayskip=\skip0
  \align}
 {\endalign\ignorespacesafterend}
 
\allowdisplaybreaks

\begin{document}
\title{Energy-Efficient mm-Wave Backhauling via Frame Aggregation in Wide Area Networks}
\author{Varun~Amar~Reddy,~\IEEEmembership{Graduate Student Member,~IEEE,}~Gordon~L.~St{\"u}ber,~\IEEEmembership{Fellow,~IEEE,}~Suhail~Al-Dharrab,\\~\IEEEmembership{Senior Member,~IEEE,}~Wessam~Mesbah,~\IEEEmembership{Senior Member,~IEEE,}~Ali~Hussein~Muqaibel,~\IEEEmembership{Senior Member,~IEEE}% <-this % stops a space
\thanks{This work is supported by the Center for Energy and Geo Processing at Georgia Institute of Technology and King Fahd University of Petroleum and Minerals (KFUPM), under research grant number GTEC1601.}
\thanks{V. A. Reddy and G. L. St{\"u}ber are with the School of Electrical and Computer Engineering, Georgia Institute of Technology, Atlanta, GA 30332, USA (e-mail: varun.reddy@gatech.edu; stuber@ece.gatech.edu).}% <-this % stops a space
\thanks{S.~Al-Dharrab, W.~Mesbah, and A.~H.~Muqaibel are with the Electrical Engineering Department, King Fahd University of Petroleum and Minerals, Dhahran 31261, Saudi Arabia (e-mail: suhaild@kfupm.edu.sa; mesbahw@gmail.com; muqaibel@kfupm.edu.sa).}% <-this % stops a space
}

\markboth{ACCEPTED BY IEEE TRANSACTIONS ON WIRELESS COMMUNICATIONS}{Reddy \MakeLowercase{\textit{et al.}}: Energy-Efficient mm-Wave Backhauling via Frame Aggregation in Wide Area Networks}

\maketitle

\begin{abstract}
Wide area networks for surveying applications, such as seismic acquisition, have been witnessing a significant increase in node density and area, where large amounts of data have to be transferred in real-time. While cables can meet these requirements, they account for a majority of the equipment weight, maintenance, and labor costs. A novel wireless network architecture, compliant with the IEEE 802.11ad standard, is proposed for establishing scalable, energy-efficient, and gigabit-rate backhaul across very large areas. Statistical path-loss and line-of-sight models are derived using real-world topographic data in well-known seismic regions. Additionally, a cross-layer analytical model is derived for 802.11 systems that can characterize the overall latency and power consumption under the impact of co-channel interference. On the basis of these models, a Frame Aggregation Power-Saving Backhaul (FA-PSB) scheme is proposed for near-optimal power conservation under a latency constraint, through a duty-cycled approach. A performance evaluation with respect to the survey size and data generation rate reveals that the proposed architecture and the FA-PSB scheme can support real-time acquisition in large-scale high-density scenarios while operating with minimal power consumption, thereby enhancing the lifetime of wireless seismic surveys. The FA-PSB scheme can be applied to cellular backhaul and sensor networks as well.
\end{abstract}
%\IEEEpeerreviewmaketitle
\section{Introduction}
\label{section:intro}
Wide area data acquisition systems for surveying applications continue to grow in area, node density, and data traffic. In particular, one such challenging application is the operation of seismic surveys for imaging the subsurface layers of the Earth to determine the location and size of oil, gas, and other mineral deposits. Seismic waves are generated by an energy source that are reflected by the subsurface layers, and in turn recorded by devices called \textit{geophones} which are deployed across the survey area. The data is then accumulated at a \textit{Data Collection Center (DCC)} and processed to generate a visual image of the Earth's subsurface. 
\par 
The subject of 2-D and 3-D seismic survey design has been well studied in \cite{Vermeer,landseismic}. A typical survey can deploy 10,000--60,000 geophones over an area of 50--300~km${}^{2}$. Given the sheer size of a seismic survey, the use of cable to connect all the geophones accounts for a majority of the equipment weight and cost. Although cable can offer high data rates in a reliable manner, a significant amount of time is spent in troubleshooting problems pertaining to the cables and connectors. Deploying cable in undulated terrain is a challenging task and can pose safety hazards to the crew. Cables can also directly impact the ecosystem of the region. For instance, it has been reported that some length of the cables is chewed upon by animals overnight~\cite{Geoexpro} leading to unmonitored ecological effects in addition to maintenance setbacks in the surveying process. \par 
While wireless systems offer an excellent alternative to cable, they come with the challenging task of achieving high data rates and relaying time-sensitive information over several nodes deployed across a widespread area. Real-time acquisition at the DCC is of vital importance as it enables field engineers to adaptively modify the acquisition parameters and minimize logistical costs. Given a seismic wavefield sampling time of 0.5~ms, a three-component (3-C) geophone with a 24-bit analog-to-digital converter would generate data at a rate of 144~Kbps. For a mid-sized survey comprising (3-C) 14,400 geophones, the data rate requirement at the DCC can reach up to 1~Gbps. Larger surveys with higher sampling rates at the geophones would mandate a requirement of nearly 5~Gbps at the DCC~\cite{Sav-1,Sav-2}. In this regard, geophone networks vary vastly from typical sensor networks where the peak data rate is on the order of a few Kbps. Additionally, power saving schemes must be incorporated since seismic acquisition can be conducted for durations lasting up to thirty days. 
\subsection{Related Work}
\label{section:related}
Several works have investigated the use of wireless systems for seismic data acquisition, where the common approach has been to deploy gateway nodes across the survey area, with each gateway node acquiring data from the geophones in its vicinity and subsequently relaying it towards the DCC. In~\cite{Sav-1}, Savazzi et al. overviewed the viability of using Bluetooth, Ultra-Wideband (UWB), WiFi, and WiMAX. The UWB-based architecture is elaborated in~\cite{Sav-2}, wherein data is transferred from geophones to nearby cluster heads, from where it is relayed to a gateway node and subsequently to the DCC. Power saving schemes have also been proposed via the use of low-power UWB radios in addition to some concepts derived from the IEEE 802.15.4 and ECMA-368 standards. Naturally, the architecture in~\cite{Sav-2} warrants tailor-made hardware with significant modifications to the ECMA framing structure. In~\cite{Patent}, seismic data is relayed serially along a chain of geophones using frequency division multiplexing in the 2.4~GHz unlicensed ISM bands. The same frequency bands are considered in~\cite{Patent2}, where relay nodes acquire data from a cluster of geophones using TDMA, which is subsequently transferred to the DCC across a mesh network of the relay nodes using full-duplex radios. However, the acquisition time may be high for data-intensive geophone networks and the aspect of power conservation has not been studied in~\cite{Patent,Patent2}. In~\cite{Compression}, a cross-layer design involving the routing policy, data compression, and resource allocation via time division multiple access (TDMA) in interference-limited regions is analyzed for a multi-hop network. A convex optimization problem is set up with the objective of minimizing the energy consumption. Although a rigorous analysis is conducted, it is done so for low data rate requirements along with the need for custom-made hardware, wherein physical (PHY) layer parameters such as the interleaving factor are also modified. An experimental study using Long Range (LoRa) technology was carried out in~\cite{lpwan}. An energy-efficient solution is provided for low-rate seismic quality control, but is not directed towards data-intensive seismic acquisition. A scalable architecture based on the IEEE 802.11af standard is described in~\cite{ICC,TWC} that can achieve power conservation in a uniform manner with the use of a minimal number of gateway nodes. Power saving techniques have also been considered in~\cite{VTC}, in which a self-organizing ad hoc network based on the IEEE 802.11ad standard is employed. The performance with respect to the latency at the DCC is evaluated in~\cite{WCNC}, for a variety of several communication technologies. \par 
Several additional approaches, outside the domain of wireless geophone networks, are of considerable interest as well. For instance, 4G and 5G networks are attractive options but their licensed nature and lack of available channel bandwidth limit their use in seismic acquisition. Low power wide area networks and \textit{MulteFire}~\cite{Multefire}, a standalone unlicensed version of LTE, offer promising solutions in terms of the power consumption and range, but cannot support data rates on the order of several gigabits per second. Some additional works have considered the use of unlicensed mm-Wave 60 GHz bands for supporting gigabit-rate energy-efficient backhaul between the 5G core network and small cells. In~\cite{backhaul1}, a TDMA-based scheduling scheme is proposed for the IEEE 802.15.3c standard. An energy-efficient user association scheme is proposed in~\cite{backhaul2} for a multi-hop backhaul network. Link scheduling schemes based on spatial TDMA have been considered in~\cite{backhaul3}. A performance comparison in terms of the energy-efficiency has been analyzed in~\cite{backhaul4} between mm-wave and optical fiber backhaul systems. 
\par
IEEE 802.11ad is a popular standard that operates over the mm-wave 60~GHz bands, and can achieve PHY rates of up to 6~Gbps, over a channel bandwidth of 2.16~GHz, through the use of a beamforming protocol for optimizing the link budget~\cite{80211-2020}. A total of 12 modulation and coding scheme (MCS) indices are available as part of the orthogonal frequency division multiplexing (OFDM) PHY, ranging from MCS 13 (700~Mbps) to MCS 24 (6.7~Gbps). Several works have shown that 802.11ad-based gigabit-rate links can be established over distances of up to a kilometer~\cite{Verma,Dehos,Rappaport}. Various 802.11ad chipsets have also been proposed for such backhaul applications~\cite{Rappaport,power11ac11ad,Sowlati}. Furthermore, high antenna gains of up to 46~dB were achieved in~\cite{Sowlati} by exploiting the small form factor that can be realized in the 60~GHz bands.
\par
Overall, existing studies pertaining to wireless geophone networks~\cite{Sav-1,Sav-2,Compression,Patent,Patent2,lpwan,ICC,TWC,VTC} have investigated data acquisition and power conservation schemes only for the links between the geophones and gateway nodes. Some of the schemes may also be licensed in nature, rendering them inaccessible and expensive. A few works~\cite{Sav-1,Sav-2,Patent,Patent2} have suggested the use of the IEEE 802.11n/ac standards for transferring data from the gateway nodes to the DCC, but have not extensively investigated such an approach. Although a latency analysis of the links between the gateway nodes and the DCC was conducted in~\cite{WCNC} with the use of Unmanned Aerial Vehicles (UAVs) and Free Space Optical (FSO) systems, power conservation schemes were not considered. While UAVs can provide excellent coverage in areas with undulating terrain, they may introduce significant latency due to flight time across several kilometers.  FSO systems can serve as high-rate backhauls but require precise and manual beam alignment, as opposed to beamforming-based protocols for mm-wave links that can auto-align the beams to optimize the link budget. Meanwhile, the approaches described in \cite{backhaul1,backhaul2,backhaul3,backhaul4} have been designed to operate over 5G networks, that differ vastly from geophone networks with respect to the propagation environment, network topology, and traffic pattern.
\subsection{Contributions}
A novel wireless geophone network architecture is proposed in this paper for high-rate energy-efficient data transfer between the gateway nodes and the DCC, which is a crucial aspect that has not been studied in previous works. The IEEE 802.11ad standard is a suitable choice that can sustain data rates of up to 6~Gbps in the 60~GHz unlicensed bands~\cite{80211-2020} and achieve real-time acquisition at the DCC even for large-scale surveys that impose an acquisition rate requirement of up to 5 Gbps~\cite{Sav-1,Sav-2}. In our previous work~\cite{VTC20}, a duty-cycling scheme was proposed for power-saving only at the geophones that were deployed in the central region of the survey area. However, in this work, a complete latency and power consumption analysis is conducted for gateway nodes that are spread across the entire survey area. Furthermore, the analytical model presented in~\cite{VTC20} is extended in this paper to include the impact of the transmit power and MCS index at the PHY layer, frame aggregation (FA) at the medium access control (MAC) layer, and the use of transmission control protocol (TCP) at the transport layer. Multiple source nodes are considered, as opposed to a single source in~\cite{VTC20}. A Frame-Aggregation Power-Saving Backhaul (FA-PSB) scheme is proposed for near-optimal power conservation across a large-scale mesh network of gateway nodes while ensuring real-time data delivery at the sink node. A duty-cycled approach is taken wherein the aforementioned analytical models are applied in determining near-optimal values for the sleep duration and data transmission parameters across the entire network. 
\par
Although this work is primarily focused on the use of IEEE 802.11ad in seismic acquisition, the underlying concepts of this paper can be extended to various other applications, such as 5G small-cell mm-wave backhaul, wireless backhaul in heterogeneous networks (HetNets), and sensor networks based on IEEE 802.11 systems for agricultural, environmental, and industrial monitoring purposes.
\par 
In summary, the primary contributions of this paper are as follows.
\begin{itemize}
\item Design of a wireless network architecture for real-time, scalable, standards-compliant, and energy-efficient backhauling across large areas. The architecture is augmented with the proposed FA-PSB scheme that relies on a duty-cycled approach for near-optimal power conservation. 
\item Design of statistical path loss and line-of-sight (LoS) models for seismic survey areas based on ray-tracing techniques using real-world topographic data.
\item Derivation of a cross-layer analytical model, based on semi-Markov processes and queuing theory, that characterizes the data transfer time and power consumption as a function of the transmit power, MCS index, frame aggregation parameters in 802.11, along with the use of TCP.
\item Derivation of an optimization framework that yields the parameters for operation under the FA-PSB scheme. A convex mixed-integer non-linear program (MINLP) is formulated and solved to achieve minimum power consumption across the network under the impact of co-channel interference and a latency constraint.
\end{itemize}
The remainder of this paper is organized as follows. The proposed network architecture and topology are described in Section II. The statistical path loss and LoS models are discussed in Section III, followed by a detailed analysis of the FA-PSB scheme in Section IV. A performance evaluation is conducted in Section V in terms of the latency and power consumption, as a function of the survey topography, survey area size, the number of geophones, and the data generation rate at each of the geophones. The findings are then summarized in the conclusion in Section VI.
\section{Proposed Network Architecture}
Following a brief discussion of the seismic surveying methodology in Section \ref{section:seismic}, a description of the proposed geophone network architecture and topology is provided in Section \ref{section:arch}.
\begin{figure*}
\centering
\includegraphics[width=1.85\columnwidth]{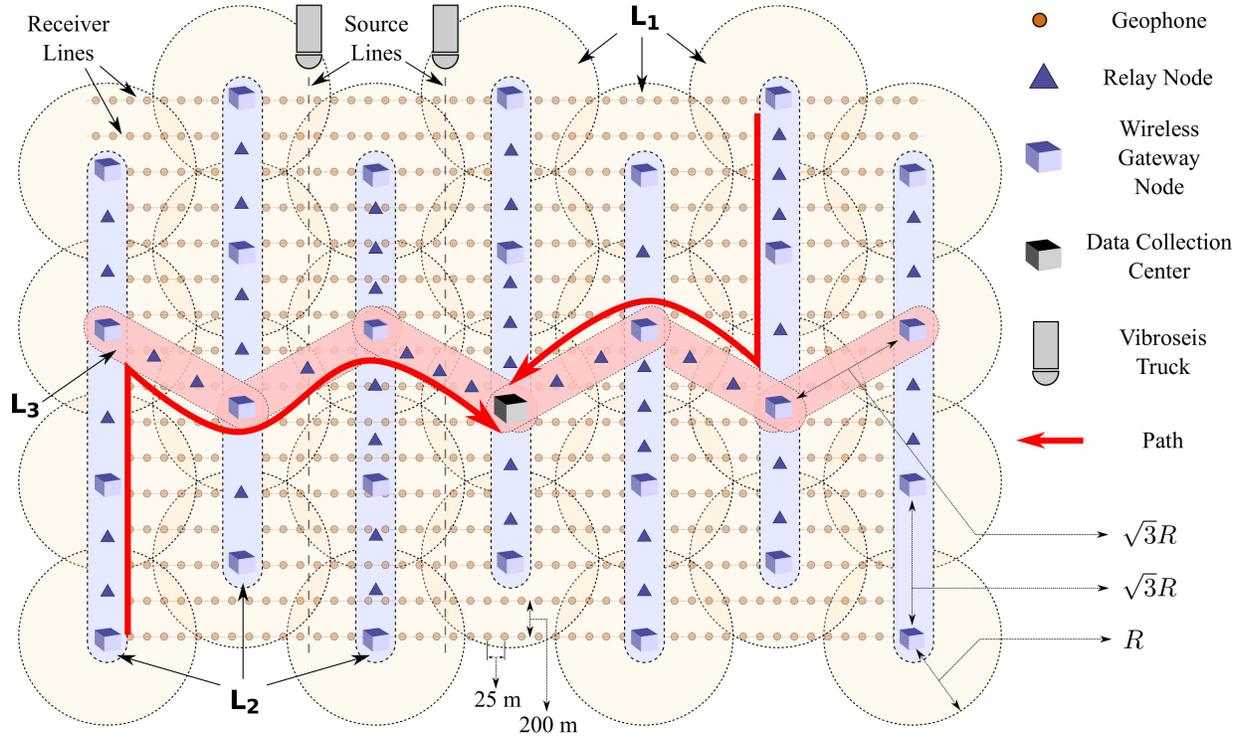}
\caption{Proposed Network Architecture} \label{fig:arch}
\end{figure*}
\subsection{Seismic Surveying Methodology}
\label{section:seismic}
An understanding of the seismic survey process is essential to determining the geophone network topology, with denser topologies resulting in superior image quality~\cite{Vermeer}. The geophones are positioned 5-30 meters apart, along a straight line to form a \textit{Receiver Line (RL)}. Vibroseis trucks move along the \textit{Source Line (SL)} and generate seismic waves, called a \textit{sweep}, for a duration of 8-12~s, known as the \textit{sweep length}. Following the sweep length, data is recorded by all the geophones for a duration of 6-8~s, known as the \textit{listen interval}. During a \textit{move-up interval} of 8-10~s, the vibroseis trucks shift to the next point where a sweep will be conducted. The three operations are repeated periodically across the survey area~\cite{landseismic}. \textit{Flip-flop} operation is considered in this study~\cite{TWC} wherein two vibroseis trucks that are sufficiently separated in distance can conduct overlapping sweeps to improve the overall productivity. Typically, there is no rigid latency requirement at the DCC. However, a good benchmark would be to acquire all the data from the previous sweep prior to the start of the listen interval of the current sweep, thereby enabling the field engineers to adaptively modify the recording parameters at the geophones. Hence, the latency threshold at the DCC can be set to the duration of the sweep length.  
\subsection{A Hierarchical Architecture}
\label{section:arch}
An orthogonal geometry is a commonly employed topology in seismic acquisition, wherein the RLs and SLs are perpendicular to one another~\cite{Vermeer}. Fig.~\ref{fig:arch} provides an illustration of the proposed network architecture that specifies an inter-geophone distance of 25~m along the RL, and an inter-RL distance of 200~m. The bottommost layer of the architecture, $\mathbf{L_{1}}$, consists of the links between \textit{wireless gateway nodes (WGNs)} and the geophones. As shown in Fig.~\ref{fig:arch}, the WGNs are laid out in a hexagonal tessellating pattern to minimize co-channel interference, where $R$ is the WGN cell radius defined as the distance from the center to the corner of the hexagon. As discussed in Section~\ref{section:related}, a variety of communication schemes have been proposed for operation at the $\mathbf{L_{1}}$ layer~\cite{Sav-1,Sav-2,Compression,Patent,Patent2,lpwan,ICC,TWC,VTC}. 
\par 
The upper $\mathbf{L_{2}}$ and $\mathbf{L_{3}}$ layers are the primary subject of interest in this paper, which are organized as a mesh network of WGNs with the DCC being the final sink node. Although both upper layers are part of the same mesh network, the notations $\mathbf{L_{2}}$ and $\mathbf{L_{3}}$ serve to demarcate the overall mesh into smaller subnets having either a vertical ($\mathbf{L_{2}}$) or a horizontal ($\mathbf{L_{3}}$) orientation. Note that the peak data transfer rates at the $\mathbf{L_{2}}$  and $\mathbf{L_{3}}$ layers are significantly higher (0.15--2.5~Gbps) as compared to the $\mathbf{L_{1}}$ layer (1.5--150~Mbps). Hence, the IEEE 802.11ad standard can be employed at the $\mathbf{L_{2}}$ and $\mathbf{L_{3}}$ layers in order to provide a real-time acquisition capability. To improve the robustness of the architecture, additional \textit{relay nodes (RNs)} can be deployed uniformly between adjacent WGNs, so as to counter large path losses in the 60~GHz bands or non-line-of-sight (NLoS) conditions. \par 
At the $\mathbf{L_{2}}$ layer, obstructions to diagonal communication links between the WGNs are inevitably created by the vibroseis trucks, thus inhibiting direct data transfer along the shortest path from the WGNs to the DCC. Tall antennas could be deployed to counter this problem, albeit requiring several such setups across the entire survey area that are robust enough to tackle high wind speeds. Hence, a more tractable solution would be to employ a directional radiation pattern at the $\mathbf{L_{2}}$ layer, so as to relay data in a direction parallel to the SLs as shown in Fig.~\ref{fig:arch}. The topmost $\mathbf{L_{3}}$ layer, comprising the DCC, forms the bottleneck of the entire network. Since the entire network topology is fixed for long durations of time, single-hop static routing can be applied. Considering a maximum antenna height of 1~m at the WGNs and RNs, it is shown in the subsequent sections of this paper that co-channel interference (CCI) between neighboring cells is effectively subdued by a larger path loss and a lower LoS probability at longer distances, implying that concurrent transmissions can occur between pairwise nodes.
\par 
A few terminologies are defined for ease of analysis in the subsequent sections of this paper. As illustrated in Fig.~\ref{fig:arch}, a \textit{path} is defined as a set of WGNs and RNs, beginning from any outer WGN at the $\mathbf{L}_{2}$ layer and leading up to the DCC. Each path comprises various \textit{links} between adjacent WGNs. Each link is further divided into \textit{sub-links} between adjacent RNs.
\begin{figure*}[t!]
        \centering
    \begin{subfigure}{\columnwidth}
        \centering
        \includegraphics[width=0.95\columnwidth]{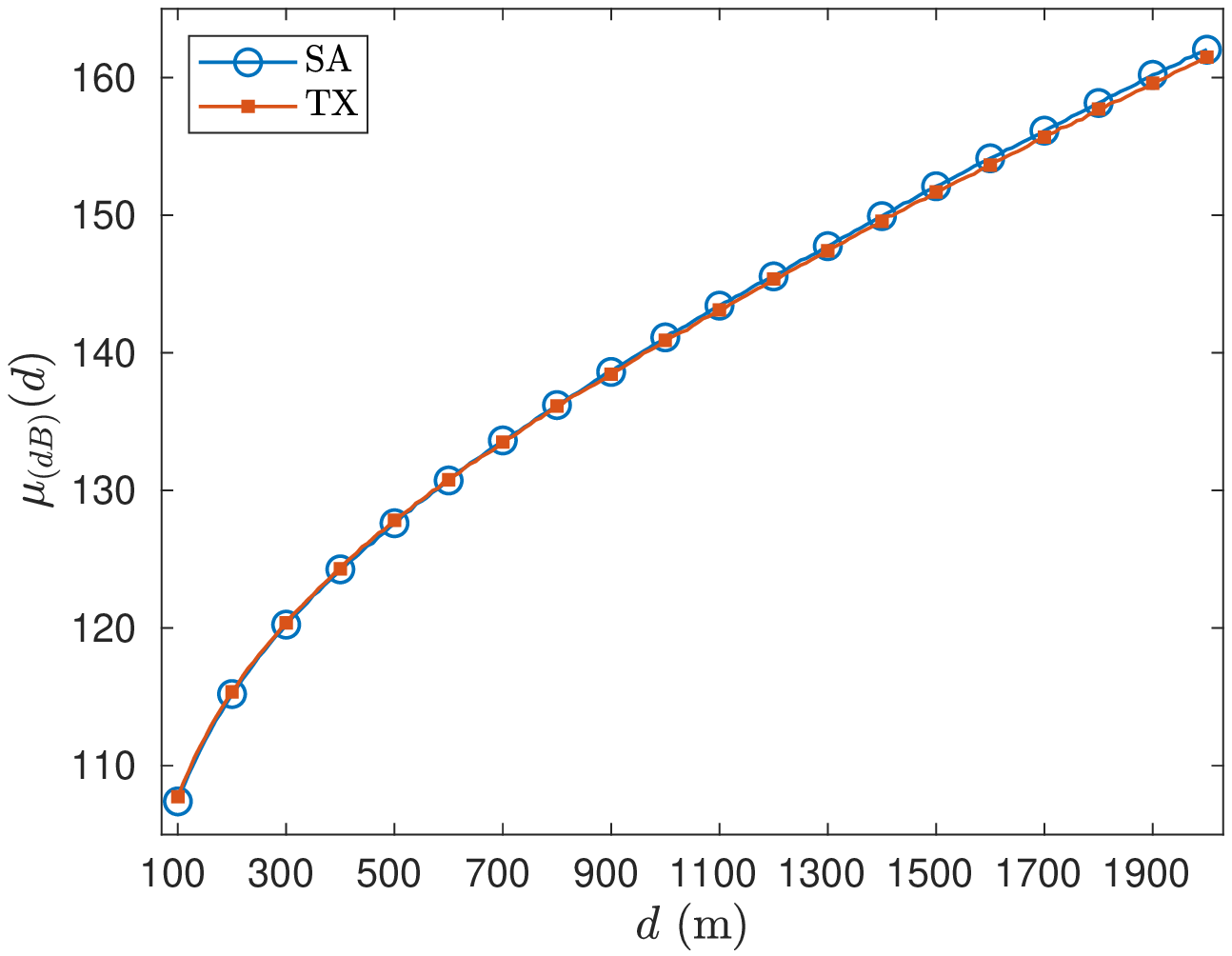}
       \caption{Mean value of the path loss.}
       \label{fig:11ad-meanloss}
    \end{subfigure} 
    \hfill
    \begin{subfigure}{\columnwidth}
        \centering
        \includegraphics[width=0.95\columnwidth]{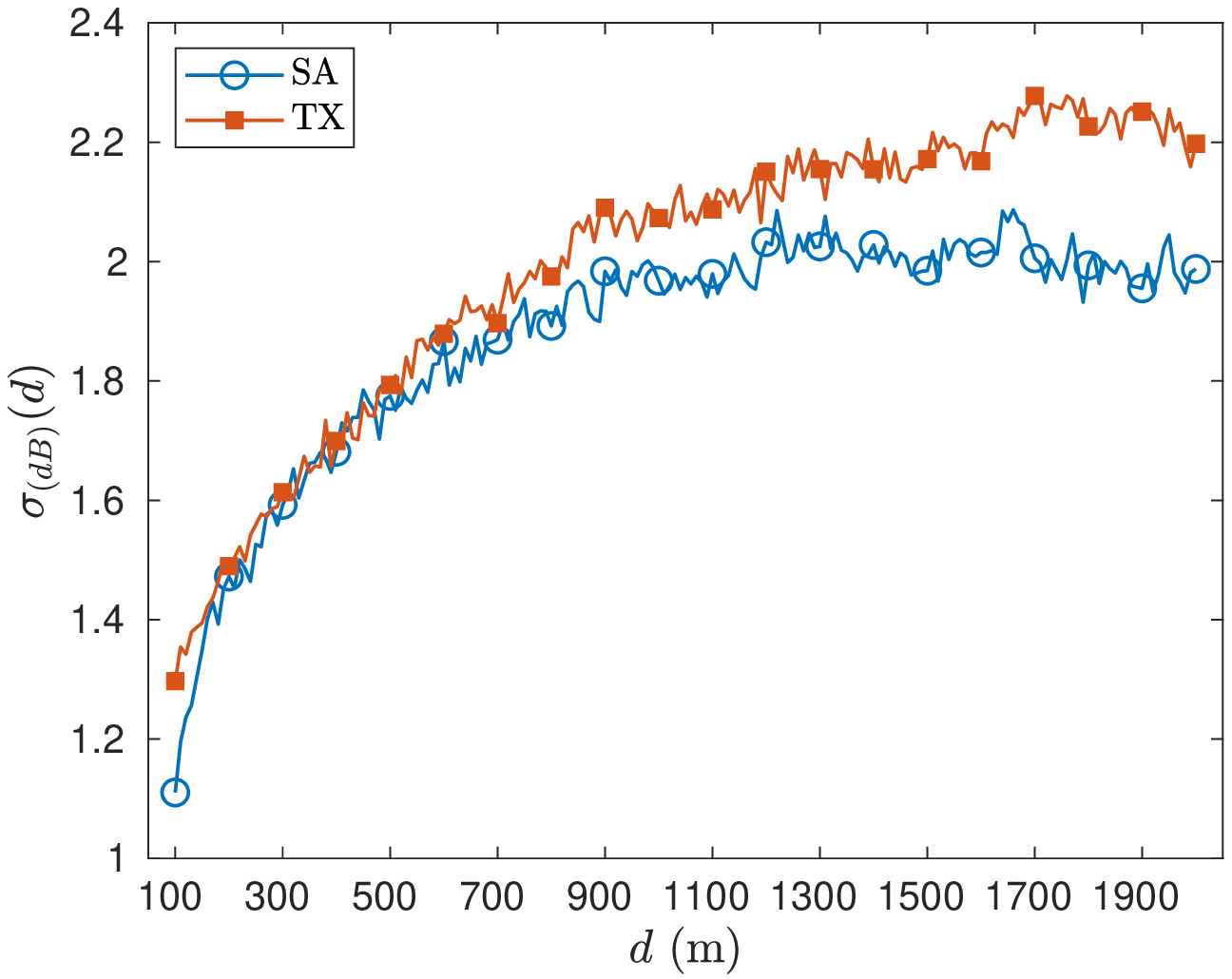}
       \caption{Standard deviation of the path loss.}
       \label{fig:11ad-stdloss}
    \end{subfigure}
   	\\[5pt]
   	    \begin{subfigure}{\columnwidth}
        \centering
        \includegraphics[width=0.95\columnwidth]{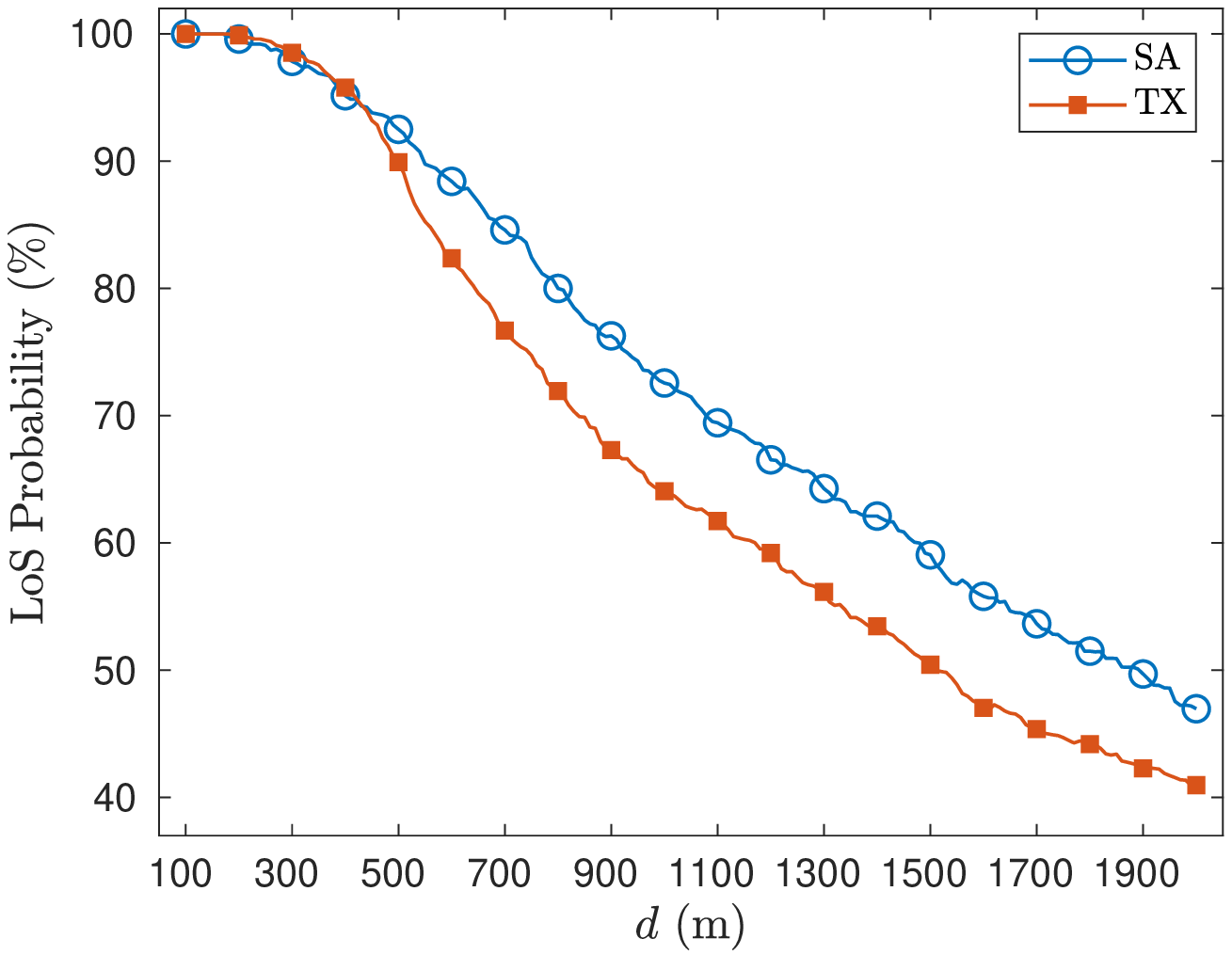}
       \caption{Line-of-sight probability as a function of the distance.}
       \label{fig:11ad-los}
    \end{subfigure} 
\caption{Statistical path loss and LoS probability models for the SA and TX surveys.} \label{fig:11ad-stats}
\end{figure*}
\section{A Statistical path loss Model for Seismic Survey Areas}
\label{section:statistical}
Modelling the propagation characteristics in seismic survey areas is vital to determining the required number of WGNs and RNs for ensuring reliable communication. In this section, statistical path loss and LoS models are derived using real-world topographic data that is available via the \textit{Open Street Maps} project~\cite{osm}. 
\par 
Let the path loss at a distance $d$ be denoted by $\text{PL}(d)$. The value of $\text{PL}(d)$ is averaged over 10,000 Monte Carlo trials, wherein two sets of geographic coordinates are randomly chosen (as per a uniform distribution) within a designated seismic survey area in each trial, such that the link distance is equal to $d$. By invoking the central limit theorem~\cite{PMC}, the effective path loss (in dB) can be treated as a log-normal random variable i.e. $\text{PL}_{(\text{dB})}(d) \sim \mathcal{N}(\mu_{(\text{dB})}(d),\sigma_{(\text{dB})}^{2}(d))$. In each trial, the total path loss is computed using a ray-tracing technique~\cite{ray-trace}, where the impact of ground reflection, surface elevation, antenna height, and atmospheric absorption loss (17~dB/km)~\cite{VTC} are taken into account. The LoS probability is also computed over the various Monte Carlo trials. \par
Considering the $l^{th}$ sub-link, let the distance, transmit power (in dB), MCS index, and the corresponding minimum required signal-to-interference-plus-noise ratio (in dB) be denoted by $d_{l}$, $\Upsilon_{l}$, $\eta_{l}$, and $\text{SINR}_{min}(\eta_{l})$ respectively. The outage probability~\cite{PMC} is given by the expression in \eqref{eq:outProb-sinr}, where the notation (dB) has been dropped to avoid overemphasis.
\begin{eqnarray}
p_{l,out}(\Upsilon_{l},\eta_{l},d_{l}) &=& \nonumber \\ && \hspace*{-3cm} Q\left(\dfrac{[\Upsilon_{l} + G_{tx} + G_{rx} - \mu(d_{l}) - \mu_{I}] - \text{SINR}_{min}(\eta_{l})}{\sqrt{\sigma^{2}(d_{l}) + \sigma_{I}^{2}}}\right) \label{eq:outProb-sinr}
\end{eqnarray}
where $G_{tx}$ and $G_{rx}$ are the \textit{realized} antenna gains at the transmit and receive sides respectively, and $Q(x)=\int_{x}^{\infty} \frac{1}{\sqrt{2\pi}}\exp(\frac{-y^{2}}{2}) dy~$ is the Gaussian Q-function~\cite{PMC}. The terms $\mu_{I}$ and $\sigma^{2}_{I}$ are the mean and variance of the interference-plus-noise power associated with the first-tier co-channel cells respectively. In the case of multiple interferers, $\mu_{I}$ and $\sigma^{2}_{I}$ can be found using a variety of techniques such as the Fenton-Wilkinson or Schwartz and Yeh methods~\cite{PMC}. 
\par 
In this work, statistical models have been generated for two well-known seismic survey projects -- the Ghawar oil field in Saudi Arabia (SA)~\cite{sa-survey} and the Permian basin in Texas (TX), USA~\cite{tx-survey}. The values for $\mu_{(\text{dB})}(d)$ and $\sigma_{(\text{dB})}(d)$ (under LoS conditions) are plotted as a function of $d$ in Fig.~\ref{fig:11ad-meanloss}-\ref{fig:11ad-stdloss}. It is seen that $\mu_{(\text{dB})}(d)$ is nearly identical for both survey areas. However, a higher value of $\sigma_{(\text{dB})}(d)$ and a lower LoS probability in the case of the TX survey suggests that it has more undulating terrain as compared to the SA survey. 
\par 
\section{Frame Aggregation Power-Saving Backhaul (FA-PSB) Scheme}
\begin{figure}
\centering
\includegraphics[width=\columnwidth]{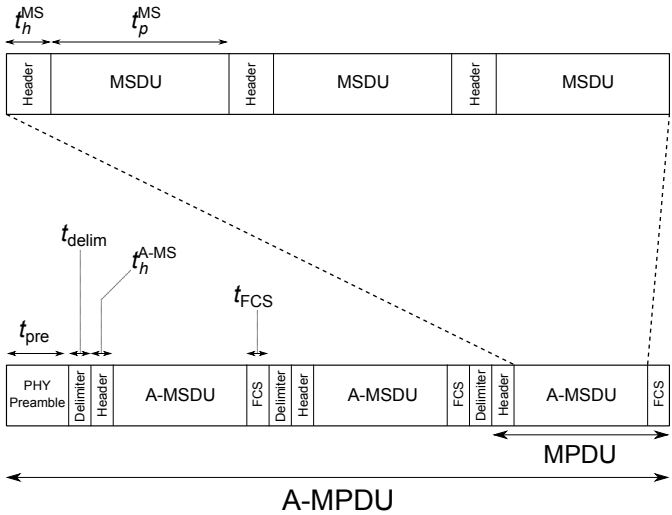}
\caption{Frame aggregation based on a combination of A-MSDU and A-MPDU~\cite{Perahia}.} \label{fig:ampdu}
\end{figure}

In addition to achieving real-time acquisition, power conservation is of vital importance as well. In the case of IEEE 802.11ad, the payload transmission time is significantly lesser as compared to the overhead time associated with the PHY header, inter-frame spaces, and MAC-layer backoff. For each data packet transmission, the overhead inadvertantly leads to an increase in the power consumption. In order to counter the impact of this overhead, \textit{frame aggregation (FA)} techniques can be employed~\cite{Perahia}. Multiple data packets can be aggregated into a single frame for transmission, thereby eliminating the recurrence of the overhead. The degree up to which FA is applied, i.e., the number of individual data packets that are aggregated, can be termed as the \textit{aggregation length}. At the MAC layer, data is encapsulated into separate MAC Service Data Units (MSDUs), which can be aggregated to form an Aggregate MSDU (A-MSDU) frame. As shown in Fig.~\ref{fig:ampdu}, the incoming MSDUs may be sequentially aggregated until a maximum size threshold is attained. An A-MSDU is then appended with a header and a frame check sequence (FCS) to form a MAC Protocol Data Unit (MPDU). As per the chosen aggregation length, the requisite number of MPDUs are in turn aggregated into an Aggregate MAC Protocol Data Unit (A-MPDU) frame. The A-MPDU frame is then passed in the egress direction to the PHY layer to form the final payload frame for transmission. A combination of both A-MSDU and A-MPDU frames has been shown to perform best~\cite{Perahia}.
\par
The performance of FA can be further improved with \textit{block acknowledgements (BAs)}~\cite{Perahia}, where the acknowledgements for each of the MPDUs can be combined into a single frame. A block acknowledgement request (BAR) is sent by the transmitter after the A-MPDU frame, following which a BA is sent back by the receiver, which contains a bitmap corresponding to the MPDUs that have failed reception. Since each MPDU is associated with its own unique FCS for error detection, only those MPDUs that were not delivered successfully are required to be retransmitted. The overall effect is a substantial reduction in the amount of overhead that would otherwise be amplified in the case of individual data-acknowledgement exchanges.
\par 
Previous studies have analyzed the use of FA for improving the channel utilization and conservation of power. In~\cite{Kuppa}, it is inferred that a dynamic assignment of the aggregation length is required as per the operating conditions of the network. An analytical model is provided in~\cite{Liu} for birectional FA using TCP. In~\cite{Assasa}, an FA-based scheduler is designed to improve the throughput in infrastructure-mode 802.11ad networks. However, the aspect of power conservation has not been considered in the above studies. FA is employed in~\cite{Adnan,Sammour} to conserve power by reducing the period of idle listening, albeit only for stations that receive frames which are not intended for them. Analytical expressions for the latency and energy consumption have been derived in~\cite{APT} for a sensor network employing the S-MAC protocol with frame aggregation. However, techniques for reducing the energy consumption were not investigated in \cite{APT}. 
\par 
The key features and operation of the proposed FA-PSB scheme are described below, following which an analytical model is developed to characterize the latency and the power consumption in Sections~\ref{fa-analysis-1} and \ref{fa-analysis-2}.
\subsection{Key Features}
\label{section:fa-kf}
\subsubsection{Effective Power Conservation} In addition to reducing the overhead with respect to time, FA can be exploited to achieve phenomenal power saving. For instance, a geophone can abstain from transmitting data for a duration of sleep in order to buffer a requisite number of packets while conserving power, after which real-time acquisition can still be perceived through a burst transmission of the buffered data using FA. The FA-PSB scheme relies on a cross-layer analytical model and optimization framework to determine the sleep duration and other parameters for data transmission, such as the transmit power, MCS index, and the aggregation length. Overall, highly effective power conservation is achieved across the backhaul network while adhering to any latency constraints at the DCC.
\par 
IEEE 802.11 devices typically require a minimum duration (denoted by $t_{sl}^{min}$) of 250~$\mu$s~\cite{Palacios} to `wake-up' from sleep mode operation. Additionally, time synchronization would have to be maintained only between adjacent nodes, where the standard-prescribed \textit{timing synchronization function} can provide an accuracy of 4~$\mu$s~\cite{TSF}, which is neglible in comparison to the value of $t_{sl}^{min}$. Hence, $t_{sl}^{min}$ can implicitly serve as a guard time for countering the possible effects of incorrect synchronization. 
\begin{figure}
\centering
\includegraphics[width=\columnwidth]{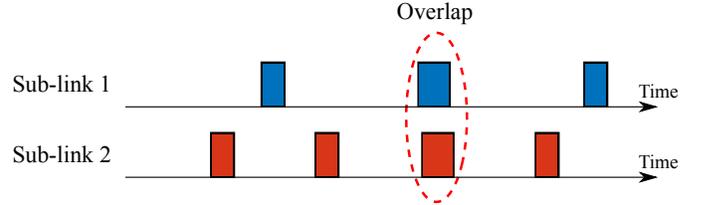}
\caption{An overlap in the data transmission periods that leads to co-channel interference between two sub-links.} \label{fig:cci}
\end{figure}
\subsubsection{Robustness against CCI} Co-channel interference can lead to a decrease in the SINR at the receiver, which in turn can lead to a higher packet error rate. In general, the impact of CCI in the proposed network is not severe, owing to a lower LoS probability and a higher path loss (due to atmospheric absorption) at larger distances. However, the FA-PSB scheme can operate in the presence of CCI as well. Given that the arrival rate of data at a node is deterministic, the resultant durations for operating in sleep mode and data transfer can be computed, following which the duty cycle for the corresponding sub-link can be found. As shown in Fig.~\ref{fig:cci}, interference would occur between two co-channel cells when the data transmission periods (represented by the `on' state of the duty cycles) overlap. Since the occurrence of such overlapping transmissions can be found in a deterministic manner~\cite{overlapProb}, the robustness of the sub-links can be ensured by altering the operating parameters such as the transmit power or the MCS index. These parameters can be obtained a-priori through a heuristic algorithm based on a combination of power and rate control that maintains a low outage probability. Hence, the deterministic nature of traffic in geophone networks can be exploited to compute the time instances when CCI would be present, following which the operating parameters are preemptively modified to ensure robustness against CCI.
\subsubsection{TCP over Mesh Networks with Large Hop-Count} It is known that TCP is not well-suited for mesh networks with a large number of hops~\cite{TCP}, since an acknowledgement from the receiver that is delayed extensively would be interpreted as packet loss by the transmitter. This problem can be circumvented by maintaining single-hop TCP links between adjacent RNs (and WGNs), rather than having a dedicated TCP connection between each of the WGNs and the DCC.
\subsubsection{Standards-Compliance} The proposed FA-PSB scheme is designed to be compliant with the TCP/IP protocol suite at the transport and network layers, along with the IEEE 802.11 protocol at the MAC and PHY layers. The functionality of the proposed FA-PSB scheme can be implemented by making appropriate changes to the device drivers or firmware, without requiring any modifications to the specifications dictated by the relevant standards. 
\begin{figure}
\centering
\includegraphics[width=\columnwidth]{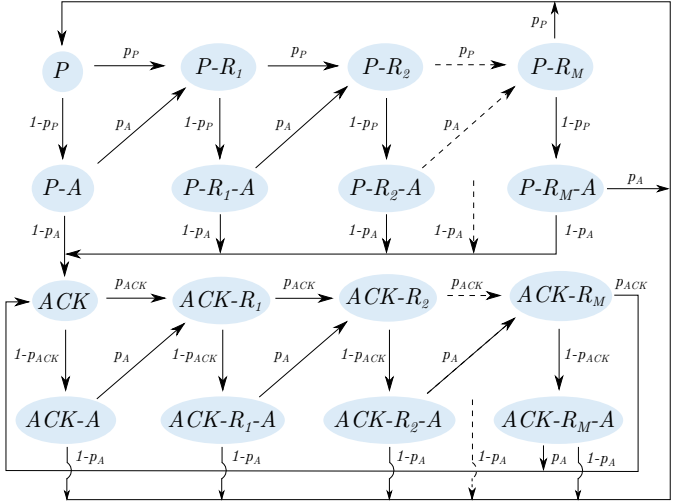}
\caption{A semi-Markov process representing TCP data transfer over the IEEE 802.11 EDCA with frame aggregation at the MAC layer.}
\label{fig:markov}
\end{figure}
\subsection{Operation}
The operation of the FA-PSB scheme is described as follows.
\begin{enumerate}[leftmargin=*]
\item For each sub-link, near-optimal values of the transmit power, MCS index, and the aggregation length are computed as a function of the latency requirement at the DCC, and the data generation rate at the trailing WGN associated with the sub-link.
\item A duty-cycled approach is employed wherein both the transmit and receive nodes operate in sleep mode for a predetermined duration, following which the buffered data is transferred using a combination of FA and BA. 
\item Transmissions can occur concurrently between adjacent sub-links over unique channels. However, in the presence of CCI, the aforementioned parameters in step 1 may be preemptively altered to ensure reliable communication.
%\item The above steps are repeated until all the data associated with numerous sweeps is delivered to the DCC.
\end{enumerate}
\subsection{Acquisition time and Power Consumption Analysis} 
\label{fa-analysis-1}
For any given sub-link, the \textit{enhanced distributed channel access} (EDCA) scheme is employed to dictate channel access, which is based on carrier sense multiple access with collision avoidance and binary exponential backoff~\cite{80211-2020}. The traffic category is considered to be best-effort, since there is no quality-of-service (QoS) differentiation between any two geophones. For each transmission, a backoff counter is introduced that is drawn uniformly from the interval $[0,CW-1]$ where $CW$ denotes the contention window size, and $CW \in [CW_{\text{min}},CW_{\text{max}}]$. The value of $CW$ is initially set to $CW_{\text{min}}$, and is doubled up to a maximum of $M$ retransmissions (when it attains the value of $CW_{\text{max}}$). After a successful transmission, $CW$ is reset back to $CW_{\text{min}}$. \par 
As shown in Fig.~\ref{fig:markov}, a semi-Markov process~\cite{Ross} is used to represent the transmission of a TCP payload segment (denoted by $P$) and up to $M$ 802.11 retransmissions (denoted by $P-R_{m},~1 \leq m \leq M$) along with the subsequent 802.11 acknowledgements (denoted by $P-A$ and $P-R_{m}-A$). This is followed by the TCP acknowledgement (denoted by $ACK$) and its associated 802.11 retransmissions and acknowledgements. The packet error probability $p_{S}$ is a function of the packet size corresponding to state $S$ and the SINR. In this analysis, the TCP congestion window is configured to match the value of the aggregation length, implying that packet collisions would not occur (since the TCP payload segment and acknowledgement would be sequentially transmitted back and forth). Hence, by assuming that there are no TCP timeouts and delayed TCP acknowledgements, the nature of $p_{S}$ is determined by the perceived SINR alone since there are no collisions that are introduced as part of the contention process between any two adjacent nodes. \par 
Considering the states related to the TCP payload segment, the value of $p_{P}$ can be found by calculating the probability that all the A-MSDUs (within the A-MPDU) and the BAR are transmitted successfully.
\begin{smalleralign}[\normalsize]
%p_{P} &= 1-(1-p_{BAR})(1-p_{BA})
p_{P} &= 1-\left\{ (1-\text{BER})^{\mathcal{S}_\text{BAR}} \times \prod\limits_{i=1}^{\mathcal{N}_\text{A-MS}} (1-\text{BER})^{\mathcal{S}_{i,\text{A-MS}}} \right\} \\
%p_{P,\text{A-MS}} &= 1-(1-\text{BER})^{\mathcal{S}_\text{A-MS}} \\
%p_\text{BAR} &= 1-(1-\text{BER})^{\mathcal{S}_\text{BAR}} \\
p_\text{A} &= 1-(1-\text{BER})^{\mathcal{S}_\text{BA}}
\end{smalleralign}
where the various notations are listed in Table~\ref{tab:80211-fa-param} and BER denotes the bit error rate. The value of $p_{ACK}$ can be found along similar lines. Given that $T_{S}$ denotes the time duration of state $S$, equations \eqref{eq:fa-time1}-\eqref{eq:fa-time5} are formulated for the $l^{th}$ sub-link operating with an aggregation length of $K_{l}$ and MCS index $\eta_{l}$. For ease of notation, define $\boldsymbol{\phi}_{l} \triangleq [\eta_{l},K_{l}]$.
\begin{smalleralign}[\normalsize]
T_{P}(\boldsymbol{\phi}_{l}) &= AIFS + (CW_{\text{min}}-1) t_{slot} /2 + t_{p}(\boldsymbol{\phi}_{l}) \label{eq:fa-time1} \\
T_{P-R_{m}}(\boldsymbol{\phi}_{l}) &= AIFS + (2^{m} CW_{\text{min}}-1) t_{slot}/2 + t_{p}(\boldsymbol{\phi}_{l}) \\
T_{ACK}(\boldsymbol{\phi}_{l}) &= AIFS + (CW_{\text{min}}-1) t_{slot} /2 + t_{ack}(\boldsymbol{\phi}_{l}) \\
T_{ACK-R_{m}}(\boldsymbol{\phi}_{l}) &= AIFS + (2^{m} CW_{\text{min}}-1) t_{slot}/2 + t_{ack}(\boldsymbol{\phi}_{l}) \\
T_{P-A}(\boldsymbol{\phi}_{l}) &= T_{ACK-A}(\boldsymbol{\phi}_{l}) = T_{P-R_{m}-A}(\boldsymbol{\phi}_{l}) \nonumber \\ &= T_{ACK-R_{m}-A}(\boldsymbol{\phi}_{l}) = SIFS + t_\text{BA}(\boldsymbol{\phi}_{l}) \label{eq:fa-time5}
\end{smalleralign}
\begin{table}[t!]
\centering
\caption{Definition of FA-specific parameters.}
\resizebox{\columnwidth}{!}{
\normalsize
\begin{tabular}
%{|c|l|}
{| C{0.1195\columnwidth} | L{0.375\columnwidth} | C{0.105\columnwidth} | L{0.405\columnwidth} |}
\hline
\textbf{Notation} & \textbf{Description} & \textbf{Notation} & \textbf{Description}\\ \hline
$\mathcal{N}_{\text{A-MS}}$ & Number of A-MSDUs in an A-MPDU & $SIFS$ & Short inter-frame space \\ \hline
$\mathcal{N}_{i,\text{MS}}$ & Number of MSDUs in the $i^{th}$ A-MSDU & $t_{slot}$ & Slot duration \\ \hline
$\mathcal{S}_{i,\text{A-MS}}$ & Number of bits in the $i^{th}$ A-MSDU & $t_{p}$ & Duration of an A-MPDU (TCP payload) \\ \hline
$\mathcal{S}_{\text{BAR}}$ & Number of bits in a BAR frame & $t_{ack}$ & Duration of an A-MPDU (TCP acknowledgement) \\ \hline
$\mathcal{S}_{\text{BA}}$ & Number of bits in a BA frame & $t_\text{BAR}$ & Duration of a BAR frame \\ \hline
$AIFS$ & Arbitration inter-frame space & $t_\text{BA}$ & Duration of a BA frame \\ \hline
\end{tabular}
\label{tab:80211-fa-param}
}
\end{table}
Referring to Fig.~\ref{fig:ampdu}, an expression for $t_{p}(\boldsymbol{\phi}_{l})$ is formulated in \eqref{eq:agg1}.
\begin{eqnarray}
\hspace*{-0.25cm} t_{p}(\boldsymbol{\phi}_{l}) &=& t_\text{BAR}(\boldsymbol{\phi}_{l}) + \nonumber \\ && \hspace*{-2cm} \underbrace{ t_\text{pre} +  \sum\limits_{i=1}^{\mathcal{N}_{\text{A-MS}}}   t_\text{delim} + \underbrace{ t_{h}^\text{A-MS} + \mathcal{N}_{i,\text{MS}} \times \{t_{h}^\text{MS} + t_{p}^\text{MS}(\eta_{l})\} + t_\text{FCS} }_{\text{Duration of } i^\text{th} \text{ A-MSDU}} }_\text{Duration of entire A-MPDU} \label{eq:agg1} 
\end{eqnarray}
Let the steady state probability of state $S$ be denoted by $\kappa_{S}$. The steady state probability values pertaining to the TCP payload segment are expressed in \eqref{phiFirst}-\eqref{phiLast}. The corresponding values pertaining to the TCP acknowledgement can be obtained by swapping the subscript $P$ with $ACK$ in \eqref{phiFirst}-\eqref{phiLast}.
\begin{smalleralign}[\small]
\kappa_{P} &= \nonumber \\ & \hspace*{-5mm} \dfrac{(1-p_{ACK})}{(4 - 3 p_{ACK} - 3 p_{P} + 2 p_{ACK} \cdot p_{P}) \sum\limits_{m=0}^{M} (p_{P} + p_{A} - p_{P} \cdot p_{A})^{m} } \label{phiFirst}
\end{smalleralign}
\begin{smalleralign}[\normalsize]
\kappa_{P-R_{m}} &= \kappa_{P} \cdot (p_{P} + p_{A} - p_{P} \cdot p_{A})^{m} \\
\kappa_{P-A} &= \kappa_{P} \cdot (1-p_{P}) \\
\kappa_{P-R_{m}-A} &= \kappa_{P} \cdot (1-p_{P}) \cdot (p_{P} + p_{A} - p_{P} \cdot  p_{A})^{m} \label{phiLast}
\end{smalleralign} 
Let $\pi_{S}$ denote the proportion of time that the semi-Markov process spends in state $S$. The value of $\pi_{S}$ can be expressed as the weighted average of the steady-state probabilities, where the weights are given by the duration spent in each state~\cite{Ross}. Define the set of states pertaining to the TCP payload segment as $S_{P} \triangleq \{P,P-R_{m},P-A,P-R_{m}-A\}$ where $m=1,2,\cdots,M$. Then, $\pi_{s} = \kappa_{s} T_{s} ~/ \sum\limits_{s' \in S_{P}} \kappa_{s'} T_{s'},~s \in S_{P}$. Similar expressions can be obtained for the states pertaining to the TCP acknowledgement, by swapping the subscript $P$ with $ACK$. \par 
A formulation can now be made for $t_{l,d}(\boldsymbol{\phi}_{l})$, which denotes the time required for a single successful exchange of a TCP payload segment and its acknowledgement over the $l^{th}$ sub-link, by accounting for those instances where the TCP payload segment or TCP acknowledgement would have to be resent after $M$ retransmissions at the MAC layer.
\begin{smalleralign}[\small]
t_{l,d}(\boldsymbol{\phi}_{l}) &= t_{l,d}^{P}(\boldsymbol{\phi}_{l}) + t_{l,d}^{ACK}(\boldsymbol{\phi}_{l}) \\[5pt]
t_{l,d}^{P}(\boldsymbol{\phi}_{l}) &= \left( \dfrac{1}{1-p_{P,\text{A-MS}}} \right) \times \nonumber \\ & \hspace*{-1cm} \left( \dfrac{\pi_{P}}{T_{P}(\boldsymbol{\phi}_{l})} - \dfrac{\pi_{P-R_{M}}}{T_{P-R_{M}}(\boldsymbol{\phi}_{l})} - \dfrac{\pi_{P-R_{M}-A}}{T_{P-R_{M}-A}(\boldsymbol{\phi}_{l})} \right)^{-1} \\[5pt] 
t_{l,d}^{ACK}(\boldsymbol{\phi}_{l}) &= \left( \dfrac{1}{{1-p_{ACK,\text{A-MS}}}} \right) \times \nonumber \\ & \hspace*{-1cm} \left( \dfrac{\pi_{ACK}}{T_{ACK}(\boldsymbol{\phi}_{l})} - \dfrac{\pi_{ACK-R_{M}}}{T_{ACK-R_{M}}(\boldsymbol{\phi}_{l})} - \dfrac{\pi_{ACK-R_{M}-A}}{T_{ACK-R_{M}-A}(\boldsymbol{\phi}_{l})} \right)^{-1} 
\end{smalleralign}
where the scaling factors $1/(1-p_{P,\text{A-MS}})$ and $1/(1-p_{ACK,\text{A-MS}})$ account for the mean number of A-MSDUs that are successfully received within an A-MPDU\footnote{Given that an A-MPDU (corresponding to the TCP payload segment) comprises $\mathcal{N}_\text{A-MS}$ A-MSDUs, the successful reception of any two distinct A-MSDUs are independent events. Hence, the mean number of total A-MSDUs that are received successfully is given by $\mathcal{N}_\text{A-MS} (1-p_{P,\text{A-MS}})$.}. \par
A similar analysis is undertaken for the power consumption. The energy consumption associated with each of the states in Fig.~\ref{fig:markov} can be expressed as follows, where $tx/rx$ denotes either the transmission or reception modes respectively. The notation $\boldsymbol{\phi}_{l}$ is dropped in \eqref{eq:fa-energy1}-\eqref{eq:fa-Prx} to avoid over-emphasis.
\begin{smalleralign}[\normalsize]
E_{P,tx/rx} &= \left( AIFS + (CW_{\text{min}}-1) \cdot t_{slot} /2 \right) I_{idle} V_{s} \nonumber \\ & \hspace*{5mm} + t_{p} I_{tx/rx} V_{s} \label{eq:fa-energy1} \\
E_{P-R_{m},tx/rx} &= \left( AIFS + (2^{m} CW_{\text{min}}-1) \cdot t_{slot} /2 \right) I_{idle} V_{s} \nonumber \\ & \hspace*{5mm} + t_{p} I_{tx/rx} V_{s}  \\
E_{ACK,tx/rx} &= \left( AIFS + (CW_{\text{min}}-1)\cdot t_{slot} /2 \right) I_{idle} V_{s} \nonumber \\ & \hspace*{5mm} +  t_{ack} I_{tx/rx} V_{s}  \\
E_{ACK-R_{m},tx/rx} &= \left( AIFS + (2^{m} CW_{\text{min}}-1) \cdot t_{slot} /2 \right) I_{idle} V_{s} \nonumber \\ & \hspace*{5mm} + t_{ack} I_{tx/rx} V_{s}  \\
E_{P-A,tx/rx} &= E_{ACK-A,tx/rx} = E_{P-R_{m}-A,tx/rx} \nonumber \\ &= E_{ACK-R_{m}-A,tx/rx} \nonumber \\ &= SIFS \cdot I_{idle} \cdot V_{s}  + t_\text{BA} \cdot I_{tx/rx} \cdot V_{s} \label{eq:fa-energy5}
\end{smalleralign}
where $V_{s}$ is the supply voltage and $I_{tx},I_{rx},I_{idle},I_{sl}$ denote the value of the current in the transmit, receive, idle, and sleep modes respectively. Note that the value of the transmit current $I_{tx}$ is dependent on the transmit power. Define the set of states $S_{P} \triangleq \{P,P-R_{1},\cdots, P-R_{M}\}$ and $S_{P-A} \triangleq \{P-A,P-R_{1}-A,\cdots, P-R_{M}-A\}$. The sets $S_{ACK}$ and $S_{ACK-A}$ can be obtained in a similar manner by replacing the subscript $P$ with $ACK$. Expressions for the power consumption in transmit and receive modes are provided in \eqref{eq:fa-Ptx} and \eqref{eq:fa-Prx} respectively. 
\begin{figure*}[t!]
\begin{smalleralign}[\normalsize]
\hspace{-2.5mm} P_{tx} &=  \frac{t_{l,d}^{P}}{t_{l,d}} \left(\sum\limits_{s' \in S_{P}} \dfrac{ \pi_{s'} E_{s',tx}}{T_{s'}}   +  \sum\limits_{s' \in S_{P-A}} \dfrac{ \pi_{s'} E_{s',rx} }{T_{s'}}  \right) + \dfrac{t_{l,d}^{ACK}}{t_{l,d}} \left(\sum\limits_{s' \in S_{ACK}} \dfrac{ \pi_{s'} E_{s',rx}}{T_{s'}} +  \sum\limits_{s' \in S_{ACK-A}} \dfrac{ \pi_{s'} E_{s',tx}}{T_{s'}} \right) \label{eq:fa-Ptx} \\[3pt]
\hspace{-2.5mm} P_{rx} &=  \dfrac{t_{l,d}^{P}}{t_{l,d}} \left(\sum\limits_{s' \in S_{P}} \dfrac{ \pi_{s'} E_{s',rx}}{T_{s'}}   +  \sum\limits_{s' \in S_{P-A}} \dfrac{ \pi_{s'} E_{s',tx} }{T_{s'}}  \right) + \dfrac{t_{l,d}^{ACK}}{t_{l,d}} \left(\sum\limits_{s' \in S_{ACK}} \dfrac{ \pi_{s'} E_{s',tx}}{T_{s'}} +  \sum\limits_{s' \in S_{ACK-A}} \dfrac{ \pi_{s'} E_{s',rx}}{T_{s'}} \right) \label{eq:fa-Prx} 
\end{smalleralign}
\hrulefill
\end{figure*}
\par
For ease of notation, define $\boldsymbol{\theta}_{l} \triangleq [\Upsilon_{l},\eta_{l},K_{l}]$ where $\Upsilon_{l}$ denotes the transmit power. In the presence of CCI, the corresponding parameters are written with a superscript `$c$'. An expression for the total power consumption of the $l^{th}$ link, comprising $r_{l}$ relays, is then given by
\begin{smalleralign}[\normalsize]
P_{l} &= \sum\limits_{l'=1}^{r_{l}+1} \left[ (1-p_{l'}^{cci}) \cdot P_{l'}^{no-cci} + p_{l'}^{cci} \cdot P_{l'}^{cci} \right] \label{eq:powerLink} \\
P_{l'}^{no-cci}  &= \frac{2\gamma_{sl}(t_{l',sl}-t_{sl}^{min}) + 2 \gamma_{idle} t_{sl}^{min} + \gamma_{d}(\boldsymbol{\theta}_{l'}) t_{l',d}(\boldsymbol{\phi}_{l'})}{t_{l',sl} + t_{l',d}(\boldsymbol{\phi}_{l'})} \label{eq:powerNoCci} \\[5pt]
P_{l'}^{cci}  &= \frac{2\gamma_{sl}(t_{l',sl}^{c}-t_{sl}^{min}) + 2 \gamma_{idle} t_{sl}^{min} + \gamma_{d}(\boldsymbol{\theta}_{l'}^{c}) t_{l',d}(\boldsymbol{\phi}_{l'}^{c})}{t_{l',sl}^{c} + t_{l',d}(\boldsymbol{\phi}_{l'}^{c})} \label{eq:powerCci}
\end{smalleralign}
As derived in Appendix~\ref{appendix:cci}, $p_{l'}^{cci}$ denotes the probability of the occurrence of CCI in the $l'^{th}$ sub-link, from which the average power is expressed in \eqref{eq:powerLink}. The parameter $t_{l',sl}$ denotes the sleep duration which has a minimum required value of $t_{sl}^{min}$. Additionally, $\gamma_{sl} = I_{sl} V_{s} $, $\gamma_{idle} = I_{idle} V_{s} $, and $ \gamma_{d}(\boldsymbol{\theta}_{l'}) =  P_{tx}(\boldsymbol{\theta}_{l'}) + P_{rx}(\boldsymbol{\theta}_{l'}) $. 
\subsection{Frame Aggregation Analysis}
\label{fa-analysis-2}
In addition to the analytical model provided in the preceding subsection, the impact of the data generation rate at the WGNs and its relation to the data transfer rate are yet to be modelled. Borrowing from queuing theory literature, the terms \textit{job} and \textit{service} would translate to a \textit{frame} and its \textit{successful reception} respectively, in the context of this work. The arrival and service rate of packets over the $l^{th}$ link is denoted by $\lambda_{l}$ and $\mu_{l}$ respectively. Note that for a given link, the arrival rate remains constant across the sub-links since the RNs do not generate any new data. \par
In the proposed FA-PSB scheme, an A-MPDU is transmitted only once $K_{l'}$ packets have been buffered. Hence, the minimum \textit{batch size} of the queue is equal to $K_{l'}$, in addition to the jobs being serviced in batches of size $K_{l'}$ as well. Overall, the transmission queue at a node in the $l'^{th}$ sub-link can be represented by a D/D$^{[K_{l'}]}$/1 \textit{batch service} model\footnote{A D/D$^{[K_{l'}]}$/1 model is considered in this analysis owing to the deterministic and static nature of traffic that is generated in wireless geophone networks. Data compression would typically retain a deterministic traffic model. However, adaptive compression and related techniques can alter the resultant traffic pattern perceived by the WGNs. In this case, a more generalized G/D$^{[K_{l'}]}$/1 batch service model must be considered for analysis.}~\cite{Ross}, wherein multiple packets are transmitted in a single A-MPDU, i.e., multiple jobs are serviced simultaneously. The domain of $K_{l'}$ is defined as the set $\mathbb{K} = \{1,~2,~3,~\cdots,~K_{max}\}$, where $K_{max}$ serves as an upper bound on the aggregation length. The value of $K_{max}$ is determined by either the maximum advertized TCP receiver window size, the coherence time of the channel, or the maximum size of an A-MPDU as indicated by the standard~\cite{80211-2020}. With the help of TCP window scaling~\cite{RFC1323}, the TCP receiver window size can be increased up to $10^{9}$ bytes. Geophone networks exhibit a static multipath environment, implying that the coherence time is sufficiently longer than the A-MPDU duration. Hence, $K_{max}$ would typically be limited by the maximum size of an A-MPDU. 
\par
Consider an expression for the service rate $\mu_{l}$ in \eqref{eq:stableQueue}, where $\mu_{l}$ must be greater than $\lambda_{l}$ to maintain queue stability.
\begin{equation}
\mu_{l} = \dfrac{K_{l'}}{t_{l',sl}+t_{l',d}(\boldsymbol{\phi}_{l'})} > \lambda_{l} \label{eq:stableQueue}
\end{equation}
For any given $\boldsymbol{\theta}_{l'}$, the value of $P_{l'}^{no-cci}$ as given by \eqref{eq:powerNoCci} monotonically decreases with an increase in $t_{l',sl}$ ($\partial P_{l'}^{no-cci}/\partial t_{l',sl} < 0, ~\forall~ t_{l',sl}>0$). The same notion applies to $P_{l'}^{cci}$ as well. Hence, minimum power consumption is achieved when the sleep duration is set to the maximum possible value while ensuring queue stability.
\begin{eqnarray}
t_{l',sl}(\boldsymbol{\theta}_{l'}) &=& \dfrac{K_{l'}}{\lambda_{l}} - t_{l',d}(\boldsymbol{\phi}_{l'}) - \delta \label{eq:sleepTime} 
\end{eqnarray}
where $\delta$ is an arbitrarily small constant that is introduced in order to satisfy the condition stated in \eqref{eq:stableQueue}. In the case of 802.11 systems, $\delta$ can be set to the value of $t_{slot}$, which represents the smallest granularity in time that is allowed by the standard. For instance, $t_{slot}$ is equal to 5~$\mu$s and 9~$\mu$s in the case of 802.11ad and 802.11ac respectively~\cite{80211-2020}. 
\par 
An expression for the latency can be now be formulated. Since a D/D$^{[K_{l}]}$/1 queue is being analyzed with $\mu_{l} > \lambda_{l}$, there is no queuing delay introduced by any of the sub-links along a path. However, a finite latency exists as a result of the end-to-end transmission delay. For a given path denoted by $\mathcal{P}$, the maximum latency $\mathcal{L}_\mathcal{P}$ at the DCC is given by
\begin{eqnarray}
\mathcal{L}_\mathcal{P} &=& \sum\limits_{l \in \mathcal{P}} \sum\limits_{l'=1}^{r_{l}+1} \left( \dfrac{K_{l'}}{\lambda_{l}} -\delta \right) \label{eq:latency}
\end{eqnarray}
In the presence of CCI, the maximum latency $\mathcal{L}_\mathcal{P}^{c}$ is given by
\begin{eqnarray}
\mathcal{L}^{c}_\mathcal{P} &=& \sum\limits_{l \in \mathcal{P}} \sum\limits_{l'=1}^{r_{l}+1} \left( \dfrac{K_{l'}^{c}}{\lambda_{l}} -\delta \right) \label{eq:latency-cci}
\end{eqnarray}
\subsection{An Optimization Framework}
With the aforementioned formulations for the latency and the power consumption, an optimization problem can now be constructed with the objective of minimizing the total power consumption at the $\mathbf{L}_{2}$ and $\mathbf{L}_{3}$ layers.
\begin{smalleralign}[\normalsize]
& \hspace*{-1cm} \minimize_{[r_{1},~\boldsymbol{\theta}_{1},~\boldsymbol{\theta}_{1}^{c},~\cdots~,~r_\mathbb{L},~\boldsymbol{\theta}_\mathbb{L},~\boldsymbol{\theta}_\mathbb{L}^{c}]}  \nonumber \\ & \hspace*{1cm} \sum\limits_{l=1}^\mathbb{L} \sum\limits_{l'=1}^{r_{l}+1} \left[ (1-p_{l'}^{cci}) \cdot P_{l'}^{no-cci} + p_{l'}^{cci} \cdot P_{l'}^{cci} \right] \label{eq:obj1} \\
P_{l'}^{no-cci}  &= 2\gamma_{sl} + \dfrac{2 (\gamma_{idle}-\gamma_{sl}) t_{sl}^{min}\lambda_{l}}{K_{l'}-\lambda_{l}\delta} + \nonumber \\ & \hspace*{2.5cm} \dfrac{ (\gamma_{d}(\boldsymbol{\theta}_{l'})-2\gamma_{sl}) t_{l',d}(\boldsymbol{\phi}_{l'}) \lambda_{l}}{K_{l'}-\lambda_{l}\delta} \label{eq:powerNoCciNew} \\
P_{l'}^{cci}  &= 2\gamma_{sl} + \dfrac{2 (\gamma_{idle}-\gamma_{sl}) t_{sl}^{min}\lambda_{l}}{K_{l'}^{c}-\lambda_{l}\delta} + \nonumber \\ & \hspace*{2.5cm} \dfrac{ (\gamma_{d}(\boldsymbol{\theta}^{c}_{l'})-2\gamma_{sl}) t_{l',d}(\boldsymbol{\phi}_{l'}^{c}) \lambda_{l}}{K^{c}_{l'}-\lambda_{l}\delta} \label{eq:powerCciNew}
\end{smalleralign}
\begin{smalleralign}[\normalsize]
t_{l',sl}(\boldsymbol{\theta}_{l'}),~t_{l',sl}(\boldsymbol{\theta}_{l'}^{c}) ~&\geq~ t_{sl}^{min}  \label{eq:cons_stable} \\
p_{l',out}\left(\boldsymbol{\psi}_{l'},\dfrac{\sqrt{3}R}{r_{l}+1}\right),~p_{l',out}\left(\boldsymbol{\psi}_{l'}^{c},\dfrac{\sqrt{3}R}{r_{l}+1}\right) ~& \leq~ p_{th,out} \label{eq:cons_outage}  \\
\mathcal{L}_\mathcal{P},~\mathcal{L}^{c}_\mathcal{P} ~&\leq~ \mathcal{L}_{max} \label{eq:cons_latency} \\
\Upsilon_{min}  ~\leq~ \Upsilon_{l'},~\Upsilon_{l'}^{c} ~&\leq~ \Upsilon_{max}  \label{eq:minTxPower} \\ 
\eta_{min} ~\leq~ \eta_{l'},~\eta_{l'}^{c} ~&\leq~ \eta_{max}  \label{eq:minMcs}  \\ 
r_{l} ~& \geq~ r^{min}_{l} \label{eq:minRelays} \\
r_{l},~\Upsilon_{l'},~\eta_{l'},~K_{l'} ~& \in ~ \mathbb{Z} \label{eq:cons_integers}
\end{smalleralign}
where $\mathbb{L}$ is the total number of links in the network and the expressions in \eqref{eq:powerNoCciNew}-\eqref{eq:powerCciNew} have been obtained by substituting for the sleep duration, as given by \eqref{eq:sleepTime}, into \eqref{eq:powerNoCci}-\eqref{eq:powerCci}. Constraint \eqref{eq:cons_stable} ensures a stable queue model and a minimum sleep duration of $t_{sl}^{min}$. Constraint \eqref{eq:cons_outage} enforces a maximum outage probability of $p_{th,out}$. The time threshold for real-time acquisition is denoted by $\mathcal{L}_{max}$, implying that \eqref{eq:cons_latency} represents the latency constraint at the DCC. Constraints \eqref{eq:minTxPower}-\eqref{eq:minMcs} provide an operating region for the transmit power and the MCS index, wherein the bounds are dependent on the standard being employed. Constraint \eqref{eq:minRelays} imposes a minimum required number of relays $r^{min}_{l}$, whose analytical expression has been derived in Appendix~\ref{appendix:relays}. Lastly, constraint \eqref{eq:cons_integers} imposes integer values on all the decision variables, where $\Upsilon_{l'}$ and $\Upsilon_{l'}^{c}$ are varied in integral steps of 1~dBm. \par
Solving the above problem is a challenging task in the presence of CCI, primarily due to the interdependency between various sub-links in \eqref{eq:outProb-sinr} which in turn renders constraint \eqref{eq:cons_outage} to be non-differentiable. In order to yield a more tractable problem, an important assumption is made wherein the presence of CCI is ignored ($p_{l'}^{cci}=0$). After solving the relaxed version of the problem, a heuristic algorithm is applied that incorporates the impact of CCI. In the subsequent sections of this paper, it is shown that such an assumption is indeed valid since the overall impact of CCI on the power consumption is not significant.
\par
The above problem is now evaluated for the following two scenarios. The latency constraint in \eqref{eq:cons_latency} is relaxed in the first scenario, which offers a twofold benefit. Firstly, an analytical advantage is obtained by removing all interdependencies between the various sub-links. Secondly, the upper bound on the latency\footnote{The upper bound on the latency is obtained by setting the aggregation length to $K_{max}$ in \eqref{eq:latency} and \eqref{eq:latency-cci} for all sub-links.} may be lesser than $\mathcal{L}_{max}$, in which case the relaxed version of the problem offers a simpler formulation while retaining its validity. Contrarily, the second scenario deals with the case when the upper bound on the latency exceeds the value of $\mathcal{L}_{max}$, mandating constraint \eqref{eq:cons_latency} to be reinstated into the optimization problem in \eqref{eq:obj1}.
\subsubsection*{\underline{Scenario 1 - No constraint on the latency}} \label{section:scenario1} 
A relaxation of the above problem in terms of both the CCI and latency eliminates any interdependencies between the sub-links. Hence, the overall problem can be decomposed into $\mathbb{L}$ unique subproblems, with each subproblem pertaining to each of the links. Furthermore, a common value for $\boldsymbol{\theta}_{l}$ can be applied to all the sub-links within the $l^{th}$ link, since $\lambda_{l}$ remains constant across all the sub-links. 
\par 
For the sake of analysis, a digression is made to define an approximation function $\Phi(\boldsymbol{\theta}_{l})$ for the third term in \eqref{eq:powerNoCciNew}, which can be interpreted as the total amount of energy consumed by the successful transfer of $K_{l}$ packets. 
\begin{smalleralign}[\normalsize]
\Phi(\boldsymbol{\theta}_{l}) &= \dfrac{ (\gamma_{d}(\boldsymbol{\theta}_{l})-2\gamma_{sl}) t_{l,d}(\boldsymbol{\phi}_{l}) }{K_{l}-\lambda_{l}\delta} \label{eq:phiActual} \\ 
&\approx  \dfrac{1}{\alpha_{1}e^{\beta_{1}K_{l}}-\alpha_{2}e^{-\beta_{2}K_{l}}} \hspace*{1cm} \alpha_{1},~\beta_{1},~\alpha_{2},~\beta_{2} ~\in~ \mathbb{R}^{+} \label{eq:phiApprox}
\end{smalleralign}
For a given value of $\boldsymbol{\psi}_{l} \triangleq [\Upsilon_{l},\eta_{l}]$, $\Phi(\boldsymbol{\theta}_{l})$ can be expressed as the reciprocal of the difference of two exponential functions, which is a commonly employed numerical approximation technique~\cite{numerical}. The proposed approximation does not introduce any loss of accuracy in the computation of $P_{l}$, and primarily serves to analytically isolate and capture the effect of $K_{l}$ on the power consumption for a given value of $\boldsymbol{\psi}_{l}$. The validity of both the aforementioned relaxation of the problem (with regards to the CCI) and the use of $\Phi(\boldsymbol{\theta}_{l})$ is justified through a performance evaluation in Section~\ref{section:performance}. 
\par
Returning to the scenario under consideration, the first derivative test reveals that $\partial \Phi(\boldsymbol{\theta}_{l})/\partial K_{l} < 0,$ $\forall ~K_{l} \in \mathbb{K}$ i.e., $P_{l}$ monotonically decreases with increasing values for $K_{l}$. This can be interpreted in a qualitative sense that an increase of the aggregation length would promote power conservation by reducing the impact of the overhead and boosting the overall data rate. Therefore, for any given $\boldsymbol{\psi}_{l}$, $P_{l}$ attains the minimum value at $K_{l}=K_{max}$. Hence, the subproblem corresponding to the $l^{th}$ link can be formulated as follows.
\begin{smalleralign}[\normalsize]
[\hat{r}_{l},~\hat{\boldsymbol{\psi}}_{l}] &= \argmin_{[r_{l},~\boldsymbol{\psi}_{l}]}~ (r_{l}+1) \left\{ 2\gamma_{sl} + \dfrac{2 (\gamma_{idle}-\gamma_{sl}) t_{sl}^{min}\lambda_{l}}{K_{max}-\lambda_{l}\delta} + \right. \nonumber \\ & \hspace*{5mm} \left. \dfrac{ (\gamma_{d}(\boldsymbol{\psi}_{l},K_{max})-2\gamma_{sl}) t_{l,d}(\eta_{l},K_{max}) \lambda_{l}}{K_{max}-\lambda_{l}\delta} \right\} \label{eq:obj2} \\
t_{sl}^{min} ~&\leq~ t_{l,sl}(\boldsymbol{\psi}_{l},K_{max})   \hspace*{2.5cm} \label{eq:obj2_cons1}  \\
p_{th,out} ~& \geq~ p_{l,out}\left( \boldsymbol{\psi}_{l},\dfrac{\sqrt{3}R}{r_{l}+1} \right)   \label{eq:cons_out} 
\end{smalleralign}
Constraints \eqref{eq:minTxPower}-\eqref{eq:cons_integers} apply to the above subproblem as well. It can be seen that the sample space is relatively small, and a simple brute force search can be performed to find the globally optimal solution. For instance, in the case of 802.11ad ($\Upsilon_{min}=-10~\text{dB},~\Upsilon_{max}=5~\text{dB},~\eta_{min}=13,~\eta_{max}=24$), a mere total of up to 2000 combinations is possible. In the case of 802.11ac ($\Upsilon_{min}=-10~\text{dB},~\Upsilon_{max}=13~\text{dB},~\eta_{min}=0,~\eta_{max}=9$), there can be up to 3000 combinations. \par
\begin{algorithm}[t!]
\small
\caption{Iterative Co-Channel Interference Mitigation}
\begin{algorithmic}[1]
\STATE Define $\Omega=\{i\},~\forall i$ s.t. $p_{i,out} > p_{th,out}$ 
\FOR {$l \in \Omega$}
\WHILE {$p_{l,out} > p_{th,out}$}
\STATE $\Upsilon_{m} \gets \Upsilon_{m} - \epsilon$  \hspace*{2cm} $\triangleright$ $m$ denotes the dominant \hspace*{4.35cm} interferer w.r.t. $l$
\IF {$p_{m,out} > p_{th,out}$}
\STATE $\eta_{m} \gets \text{max}\{\eta_{m} - 1,\eta_{min}\}$  
\ENDIF
\ENDWHILE
\IF {$m \notin \Omega$}
\STATE $\Omega \gets \Omega \cup \{m\}$  \hspace*{2cm} $\triangleright$ Include $m$ in the set $\Omega$, \hspace*{4.375cm} if not present already
\ENDIF
\ENDFOR
\end{algorithmic}
\label{alg:cci}
\end{algorithm}
The impact of CCI is now taken into account via Algorithm~\ref{alg:cci} which is heuristic in nature. Considering the solution obtained by solving~\eqref{eq:obj2}, the sub-links that are severely affected by CCI are grouped into a set denoted by $\Omega$. For each of these affected sub-links, the transmit power for the dominantly-interfering sub-link can be iteratively reduced until the impact of CCI is nullified. In the event that the interfering sub-link can no longer operate at the originally desired MCS index, the MCS index is decremented as well. This procedure is repeated iteratively for all the remaining co-channel cells. Hence, the proposed heuristic algorithm can be interpreted as a combination of power and rate control, which rapidly converges to a feasible solution (owing to a small sample space). The motivation for performing power control prior to rate control arises from the fact that the power consumption is more susceptible to a decrease in the MCS index as compared to an increase in the transmit power~\cite{power11ac11ad}.  
\subsubsection*{\underline{Scenario 2 - A constraint on the latency}} During a seismic survey, the latency requirement at the DCC may be periodically altered by the field engineers. Furthermore, in the case of a low arrival rate of data (such as in earthquake detection or quality control) the upper bound on the latency would exceed $\mathcal{L}_{max}$, and it would be necessary to impose a latency constraint at the DCC. An important point to consider is that for the $l^{th}$ link, adaptively modifying $\boldsymbol{\theta}_{l'}$ for each of the sub-links within the link is a feasible option, whereas an adaptive change in the number of relays is practically infeasible. Introducing or moving existing RNs during the acquisition process over the entire survey area, can drastically reduce the overall productivity and may even be inconceivable. Hence, a more tractable approach would be to retain the values obtained for $\hat{r}_{l}$ and $\hat{\boldsymbol{\psi}}_{l}$ in the preceding scenario, and instead adaptively modify only the value of the aggregation length so as to meet the latency constraint at the DCC. The latency in the worst-case scenario is considered by treating the presence and absence of CCI as two distinct optimization problems. The following problem is formulated in the absence of CCI.  \par 
\begin{smalleralign}[\normalsize]
[\hat{K}_{1},~\hat{K}_{2}~\cdots~\hat{K}_\mathbb{L}] = \nonumber \\ & \hspace*{-3cm} \argmin_{K_{1},K_{2},~\cdots,~K_\mathbb{L}} ~\sum\limits_{l=1}^\mathbb{L} \sum\limits_{l'=1}^{r_{l}+1} 2\gamma_{sl} + \dfrac{2 (\gamma_{idle}-\gamma_{sl}) t_{sl}^{min}\lambda_{l}}{K_{l'}-\lambda_{l}\delta} \nonumber \\ & \hspace*{2cm} +  \lambda_{l} \Phi(K_{l'},\hat{\boldsymbol{\psi}}_{l'}) \label{eq:obj3} \\
\mathcal{L}_\mathcal{P} ~&\leq~ \mathcal{L}_{max} \hspace*{3cm} \label{eq:cons_maxLatency}  \\
t_{l',sl}(\hat{\boldsymbol{\psi}}_{l'},K_{l'}) ~&\geq~ t_{sl}^{min} \label{eq:cons_sleep1_obj3}
\end{smalleralign}
A latency constraint is imposed by \eqref{eq:cons_maxLatency} for all possible paths in the network. The lower bound for $K_{l}$ can be found using \eqref{eq:cons_sleep1_obj3}. Furthermore, it is proven in Appendix~\ref{appendix:convex} that $\Phi(\boldsymbol{\theta}_{l'})$ is convex $\forall ~K_{l'} \in \mathbb{K}$. The overall objective function as given in \eqref{eq:obj3} is also convex, since a linear composition of convex functions is convex~\cite{Boyd}. Hence, the above convex MINLP can be solved using standard techniques~\cite{Boyd,convexMINLP}. Note that the equivalent problem in the presence of CCI can be formulated by introducing the superscript notation $c$ for all the relevant parameters. The resultant total power consumption is then found using the expression in~\eqref{eq:obj1}.
\begin{figure*}[t!]
        \centering
    \begin{subfigure}{\columnwidth}
        \centering
        \includegraphics[width=0.95\columnwidth]{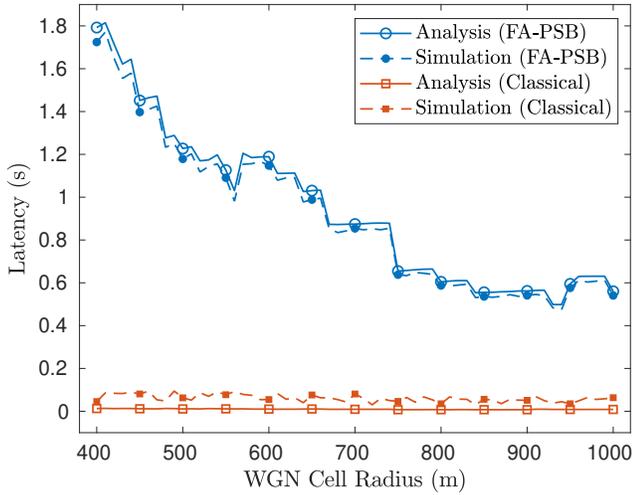}
       \caption{Latency at the DCC (SA survey).}
       \label{fig:results-1a}
    \end{subfigure} 
    \hfill
    \begin{subfigure}{\columnwidth}
        \centering
        \includegraphics[width=0.95\columnwidth]{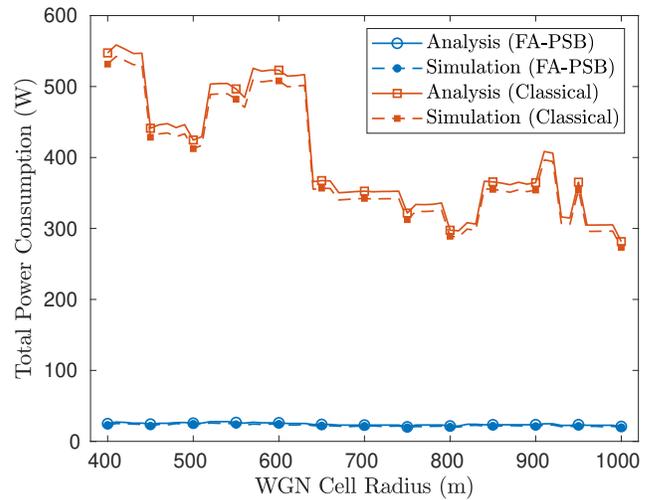}
       \caption{Total power consumption (SA survey).}
       \label{fig:results-1b}
    \end{subfigure}
   	\\[10pt]
   	    \begin{subfigure}{\columnwidth}
        \centering
        \includegraphics[width=0.95\columnwidth]{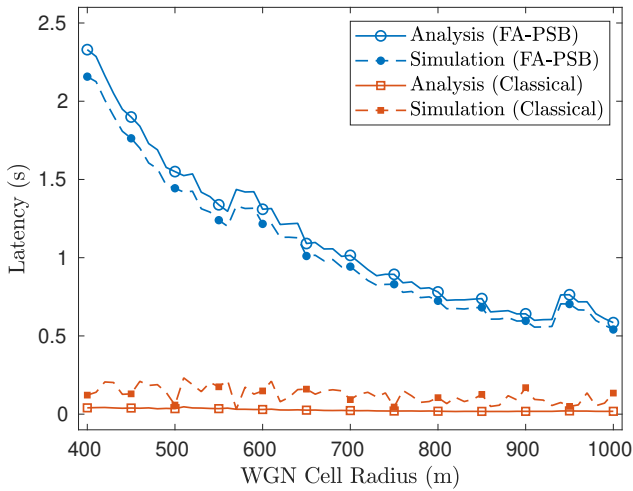}
       \caption{Latency at the DCC (TX survey).}
       \label{fig:results-1c}
    \end{subfigure} 
    \hfill
    \begin{subfigure}{\columnwidth}
        \centering
        \includegraphics[width=0.95\columnwidth]{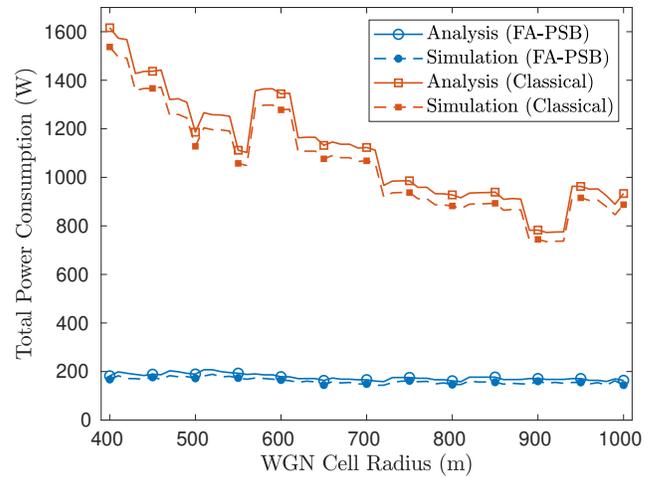}
       \caption{Total power consumption (TX survey).}
       \label{fig:results-1d}
    \end{subfigure}
\caption{A performance comparison between classical operation and the FA-PSB scheme (with no latency constraint).} \label{fig:results-1}
\end{figure*}
\section{Performance Evaluation}
\label{section:performance}
Considering a 4-cell reuse pattern and a sweep length of 8~s, the performance with respect to the latency and power consumption (at the $\mathbf{L}_{2}$ and $\mathbf{L}_{3}$ layers) is evaluated for the SA and TX survey terrain. In the following performance evaluation, the SA survey represents a mid-sized survey comprising 14,400 geophones over an area of 72~km$^{2}$, while the TX survey represents a large survey comprising 57,600 geophones over an area of 288~km$^{2}$. The resultant values are averaged over 1000 Monte Carlo trials, wherein the geophone network is deployed over a random section (as per a uniform distribution) of the seismic survey region in each trial. Simulations are conducted using \textit{ns-3} for the IEEE 802.11ad standard~\cite{ns3-80211ad}. The simulation parameters have been listed in Tables~\ref{tab:Total-Sim} and under Appendix~\ref{appendix:values}. Standards-compliant values of $CW_{\text{min}}=16$ and $CW_{\text{max}}=1024$ are considered, such that $M=6$. The parameters in Tables~\ref{tab:80211-fa-param}-\ref{tab:Total-Sim} are derived from~\cite{80211-2020,80211-survey}. Experimental values for the power consumption parameters of IEEE 802.11ad and IEEE 802.11ac chipsets have been reported in~\cite{Sowlati} and~\cite{power11ac} respectively.  \par
\begin{table}[t!]
\centering
\normalsize
\caption{Simulation parameters.}
\resizebox{\columnwidth}{!}{
\begin{tabular}{| L{0.345\columnwidth} | C{0.135\columnwidth} | L{0.365\columnwidth} | C{0.155\columnwidth} |}
\hline
\textbf{Parameter} & \textbf{Value} &
\textbf{Parameter} & \textbf{Value}\\ \hline
Operating Frequency & 57-64 GHz & Maximum size of an A-MSDU frame & 7935 bytes \\ \hline
Channel Bandwidth & 2.16 GHz & Maximum size of an A-MPDU frame & 262,143 bytes \\ \hline
Antenna Height & 1~m &  Supply Voltage $(V_{s})$ & 3~V \\ \hline
Maximum EIRP & 51~dBm & Current in transmit mode $(I_{tx})$ & 2776~mA  \\ \hline
Noise Figure & 6~dB & Current in receive mode $(I_{rx})$ & 2198~mA  \\ \hline
Outage probability threshold ($p_{th,out}$) & $10^{-6}$ & Current in idle mode $(I_{idle})$ & 420~mA \\ \hline
TCP Maximum Segment Size & 2200 bytes & Current in sleep mode $(I_{sl})$ & 5~mA \\ \hline
\end{tabular}
\label{tab:Total-Sim}
}
\end{table}
The latency and the total power consumption for the SA and TX surveys operating under the FA-PSB scheme is shown in Fig.~\ref{fig:results-1} as a function of the WGN cell radius $R$, for a geophone data generation rate of 144~Kbps. While the latency ranges between $0.5-2.2$~s, which is well within the value of $\mathcal{L}_{max}$ (the sweep length), the power saving performance is phenomenal, with a $78-87$\% drop in the total power consumption under the FA-PSB scheme as compared to the classical scenario (there is no sleep duration imposed on the sub-links). The number of WGNs decreases with increasing values of $R$~\cite{TWC}, which explains the descending trend in the latency and power consumption\footnote{As derived in~\cite{TWC}, the number of WGNs does not smoothly decrease with an increase in the value of $R$, which in turn imposes a non-monotonic decreasing behavior on both the latency and power consumption.}. However, certain non-linearities are introduced in the trend since the total number of RNs does not monotonically decrease with $R$. For instance, an abrupt increase in the latency and the power consumption is seen in both surveys around $R=570~$m. This is due to the fact that the outage probability threshold in~\eqref{eq:cons_outage} can only be met by the introduction of additional RNs. Hence, although it may seem counter-intuitive that the power consumption would increase despite a drop in the number of WGNs, it is important to consider the number of RNs that are required to sustain reliable and real-time acquisition for larger values of $R$.
\par
It can also be inferred from Fig.~\ref{fig:results-1c}-\ref{fig:results-1d} that the overall latency and power consumption in the TX survey is larger as compared to the SA survey, which arises due to two factors. Firstly, the TX survey area mandates a larger number of WGNs and RNs as it is four times larger in size. Secondly, the TX survey terrain is characterized by a lower LoS probability (as shown in Fig.~\ref{fig:11ad-los}) which in turn would require a larger number of RNs to be deployed. Furthermore, the effectiveness of the proposed analytical technique can be seen in Fig.~\ref{fig:results-1}. A small but finite margin of error exists between the analysis and simulation results, due to the assumption of perfectly saturated traffic (which in turn leads to an overestimation in the computation of the packet error probability) and the statistical characterization of the probability of the occurrence of CCI. However, the overall trend is captured well by the proposed analytical model. 
\begin{figure*}[t!]
        \centering
   	    \begin{subfigure}{\columnwidth}
        \centering
        \includegraphics[width=0.95\columnwidth]{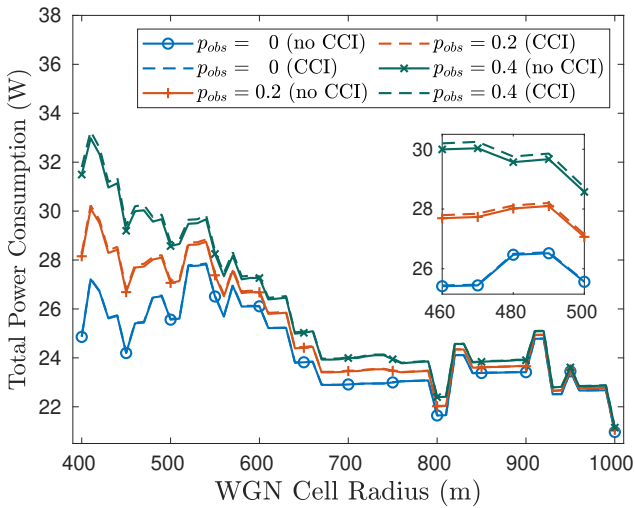}
       \caption{Total power consumption (SA survey).}
       \label{fig:results-2a}
    \end{subfigure} 
    \hfill
    \begin{subfigure}{\columnwidth}
        \centering
        \includegraphics[width=0.95\columnwidth]{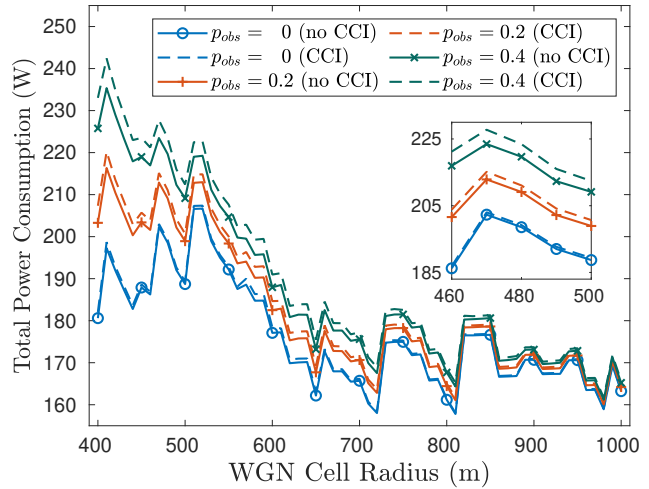}
       \caption{Total power consumption (TX survey).}
       \label{fig:results-2b}
    \end{subfigure}    
\caption{A comparison between the solutions obtained via the proposed heuristic algorithm (in the presence of CCI) and the optimal solution (in the absence of CCI) for various values of $p_{obs}$.} \label{fig:results-2}
\end{figure*}
\begin{figure}
        \centering
		\includegraphics[width=0.95\columnwidth]{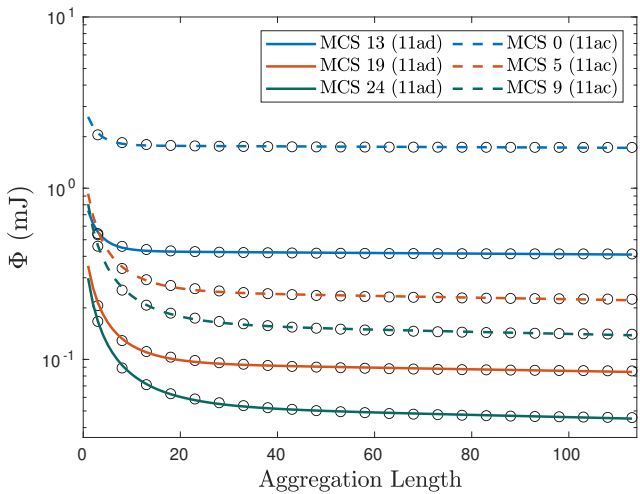}		       
        \caption{Validity of the convex approximation function $\Phi(\bullet)$.}
        \label{fig:convexFit}
\end{figure}
\begin{figure*}[tp!]
    \centering
    \begin{subfigure}{\columnwidth}
        \centering
        \includegraphics[width=0.95\columnwidth]{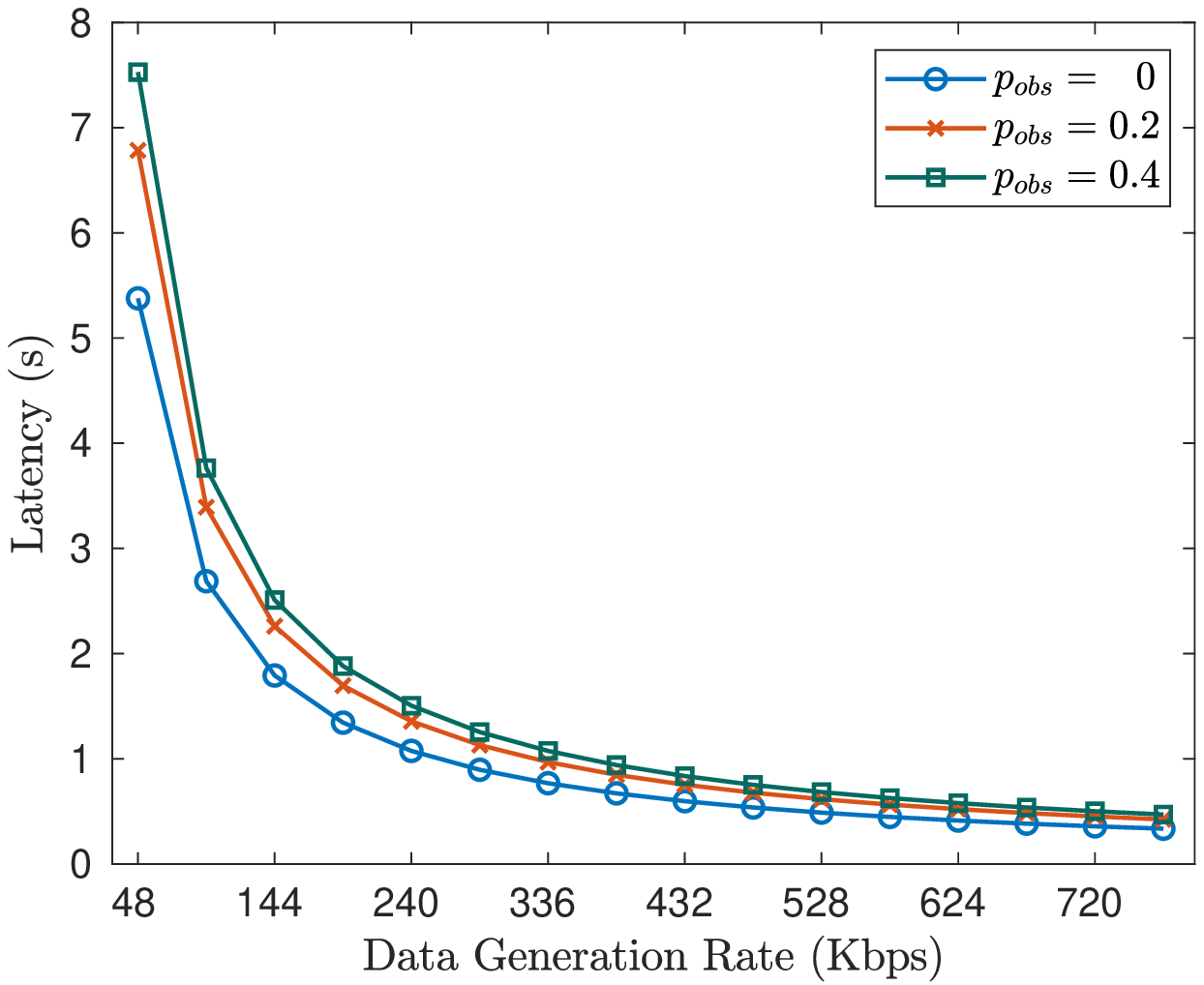}
       \caption{Latency at the DCC (SA survey).}
       \label{fig:results-3a}
    \end{subfigure} 
    \hfill
    \begin{subfigure}{\columnwidth}
        \centering
        \includegraphics[width=0.95\columnwidth]{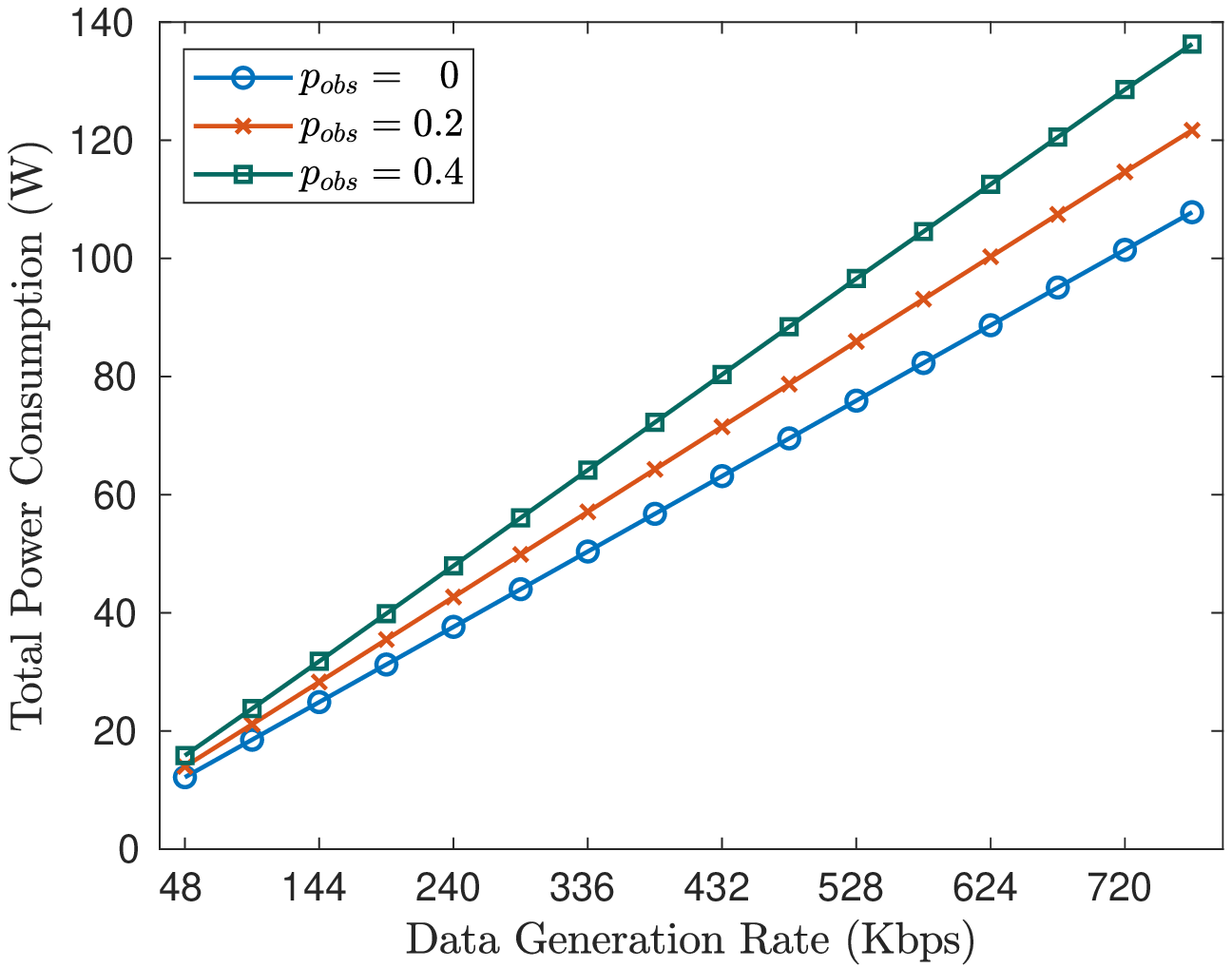}
       \caption{Total power consumption (SA survey).}
       \label{fig:results-3b}
    \end{subfigure}
   	\\[10pt]
   	    \begin{subfigure}{\columnwidth}
        \centering
        \includegraphics[width=0.95\columnwidth]{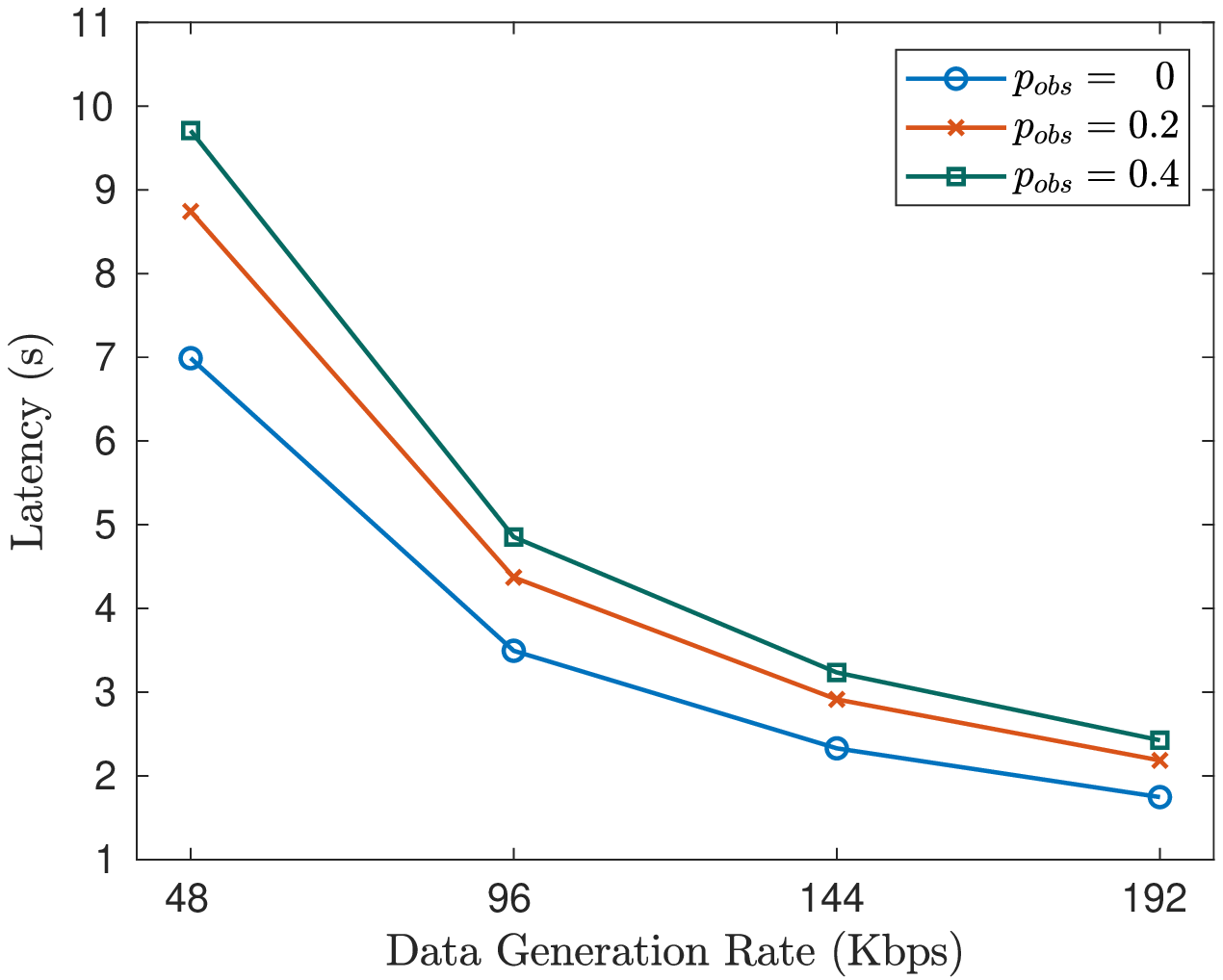}
       \caption{Latency at the DCC (TX survey).}
       \label{fig:results-3c}
    \end{subfigure} 
    \hfill
    \begin{subfigure}{\columnwidth}
        \centering
        \includegraphics[width=0.95\columnwidth]{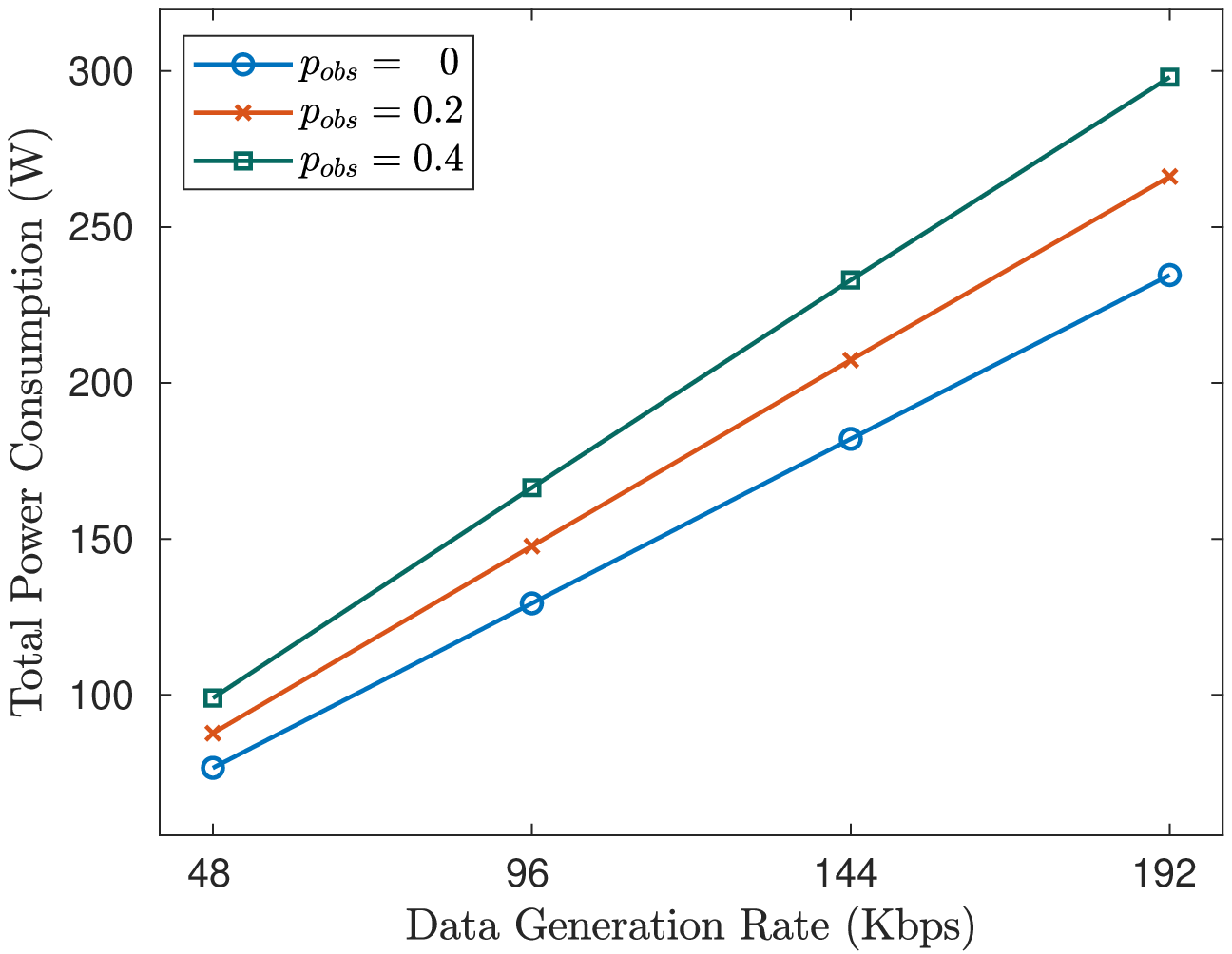}
       \caption{Total power consumption (TX survey).}
       \label{fig:results-3d}
    \end{subfigure}
\caption{Latency and power consumption performance as a function of the data generation rate and $p_{obs}$.} \label{fig:results-3}
\end{figure*}
\par
The efficacy of the proposed heuristic algorithm in Section~\ref{section:scenario1} is validated in Fig.~\ref{fig:results-2}. The additional impact of NLoS links due to the presence of obstacles is also considered, wherein $p_{obs}$ denotes the fraction of the total number of links that are obstructed. Consequently, each of these links would require an additional RN to be deployed. The proposed heuristic algorithm provides a solution in the presence of CCI (denoted by dashed lines) that is very close to the globally optimal solution in the absence of CCI (denoted by solid lines), implying that the overall impact of CCI is minimal. This arises from the fact that the co-channel cells are mostly distant from one another wherein the CCI is effectively subdued. The overall power consumption increases with $p_{obs}$ owing to an increase in the number of RNs. This in turn marginally worsens the impact of CCI wherein the co-channel cells are now located closer to one another. Additionally, the validity of the approximation function $\Phi$ is justified in Fig.~\ref{fig:convexFit}, where the circles represent the actual value as given by \eqref{eq:phiActual} and the solid and dashed lines represent the approximate value as given by \eqref{eq:phiApprox}. A maximum relative RMSE value of just 1\% is noted for the various combinations of the MCS index for a transmit power of 5~dBm in the case of both 802.11ac and 802.11ad. 
\par
In Fig.~\ref{fig:results-3}, the impact of the data generation rate at each of the geophones (multiples of 48~Kbps) is studied for $R = 400$ m. As seen in Fig.~\ref{fig:results-3a}-\ref{fig:results-3b}, a maximum rate of 768~Kbps can be sustained by the proposed 802.11ad-based architecture for the SA survey. A higher data generation rate at the geophones would cause \eqref{eq:stableQueue} and \eqref{eq:cons_stable} to not be satisfied for the bottleneck links at the $\mathbf{L}_{3}$ layer, which in turn would lead to queue instability and consequently an exponential latency at the DCC. The latency observes a decreasing trend in Fig.~\ref{fig:results-3a} and Fig.~\ref{fig:results-3c}, since the sleep duration is reduced in order to maintain queue stability at higher values of the packet arrival rate. A corresponding increase is seen in the power consumption, where the WGNs and RNs are required to operate in transmit and receive modes for a greater fraction of time. In the case of the TX survey, a maximum data generation rate of 192~Kbps can be applied at the geophones. Note that the present analysis deals with the transfer of raw data from the geophones to the DCC. However, data compression techniques can be utilized to reduce the effective packet arrival rate, which in turn can substantially enhance the power conservation performance of the FA-PSB scheme.  
\par
\begin{figure*}[t!]
    \centering
    \begin{subfigure}{0.95\columnwidth}
        \centering
        \includegraphics[width=\columnwidth]{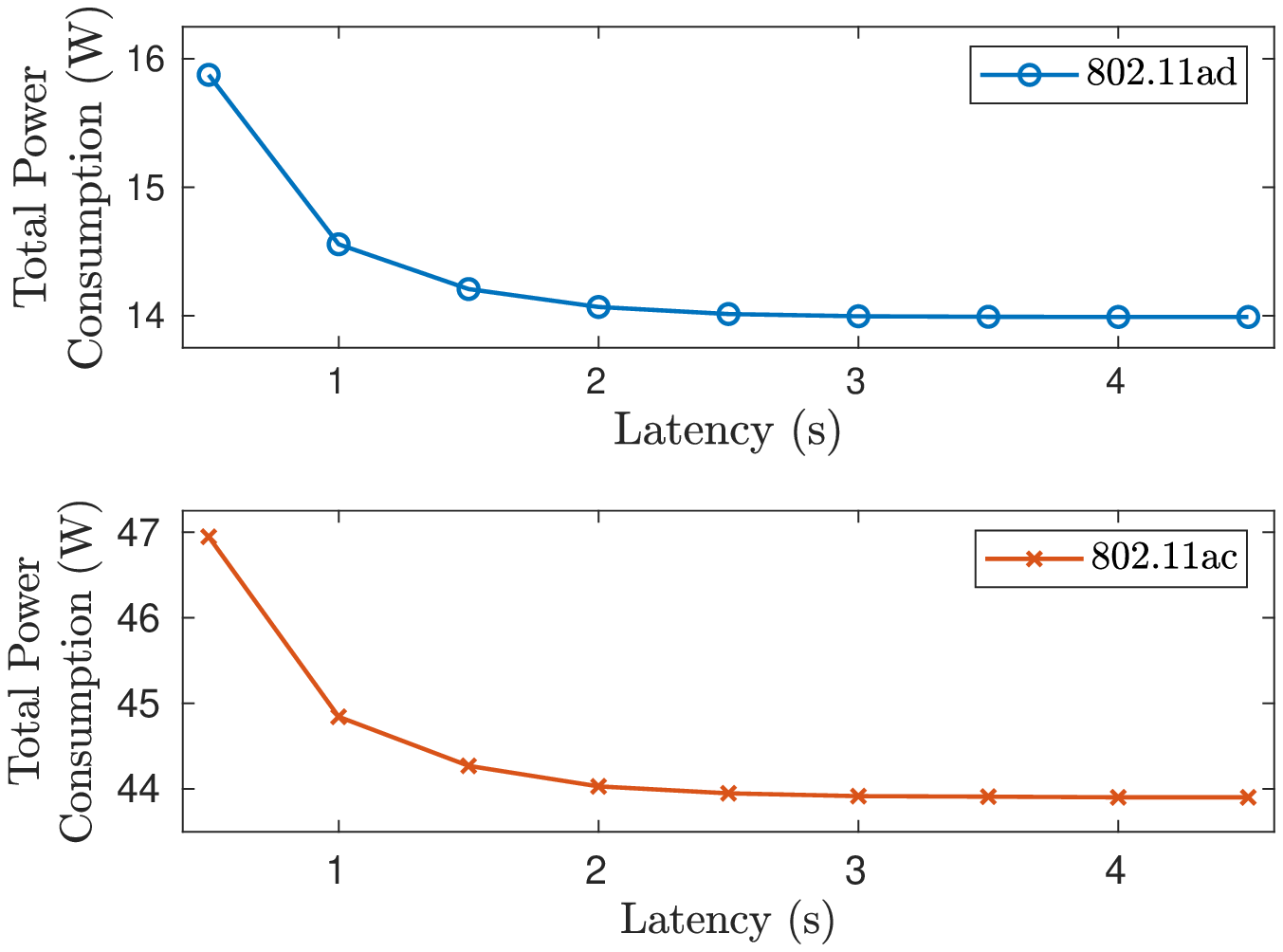}
       \caption{Evaluation at a data generation rate of 48~Kbps.}
       \label{fig:results-4a}
    \end{subfigure} 
    \hfill
    \begin{subfigure}{0.95\columnwidth}
        \centering
        \includegraphics[width=\columnwidth]{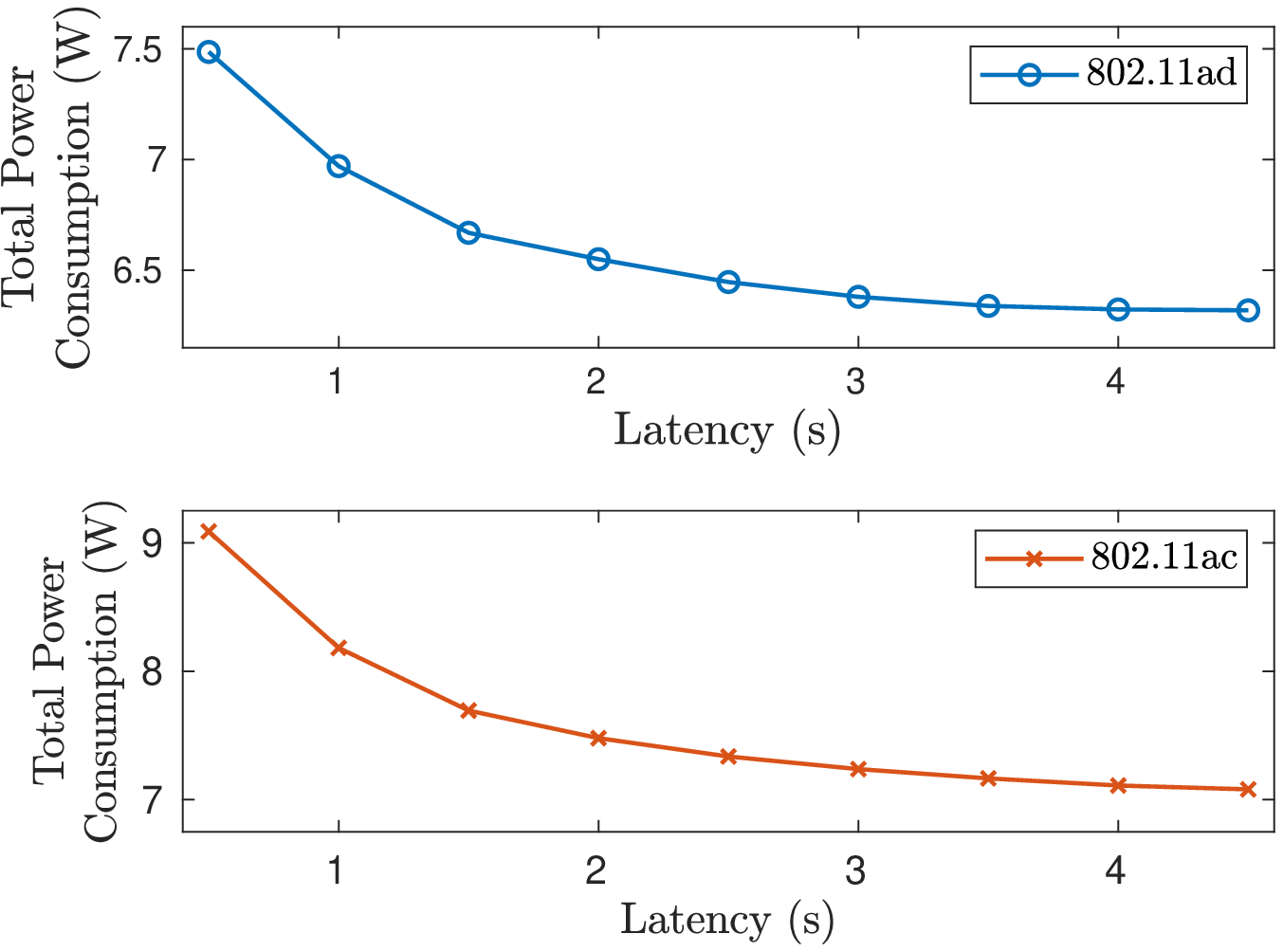}
       \caption{Evaluation at a data generation rate of 1~Kbps.}
       \label{fig:results-4b}
    \end{subfigure}
\caption{Performance comparison between the 802.11ad and 802.11ac standards in terms of the trade-off between the latency and power consumption (SA survey).} \label{fig:results-4}
\end{figure*}

A performance comparison is now conducted between the use of the IEEE 802.11ad and IEEE 802.11ac standards under a latency-constrained scenario for the SA survey. When low-rate alternative applications such as seismic quality control or earthquake detection are considered, where the data generation rate can be as low as 1~Kbps, a latency of several minutes may be introduced by the FA-PSB scheme. A latency constraint would be deemed necessary in such scenarios, at the cost of a marginal increase in the power consumption. The tradeoff between the latency and the power consumption is shown in Fig.~\ref{fig:results-4} for the SA survey and for $R = 400$ m. Considering a data generation rate of 48~Kbps in Fig.~\ref{fig:results-4a}, it can be seen that the 802.11ad standard is able to achieve a much lower power consumption as compared to 802.11ac, by exploiting its gigabit-rate capability to increase the sleep duration while maintaining queue stability. In the case of a data generation rate of 1~Kbps in Fig.~\ref{fig:results-4b}, the power conservation benefit that is provided by the 802.11ad standard is only marginal as compared to the 802.11ac standard. This arises from the fact that the value of $t_{l',d}\lambda_{l}$ in~\eqref{eq:phiActual} is effectively lowered by a very small value for $\lambda_{l}$, which in turn dwarfs the impact of a lower value of $t_{l',d}$ in the case of 802.11ad as compared to 802.11ac. Hence, the 802.11ac standard may be deemed a feasible choice in low-rate applications but fails to provide satisfactory results in the case of data-intensive seismic acquisition. For instance, the maximum data generation rates that can be sustained by the 802.11ac standard are only 1 and 48 Kbps in the case of the SA survey, and just 1 Kbps in the case of the TX survey. Since the maximum PHY-layer rate is only around 440~Mbps for the standard 80~MHz channels~\cite{80211-2020}, higher data generation rates at the geophones would lead to queue instability and an exponential latency at the DCC.
\section{Conclusion}
Seismic acquisition at the data collection center, the final sink node for the entire geophone network, mandates data transfer rates on the order of several gigabits per second in order to have real-time data delivery, in addition to the acquisition process being energy-efficient. A wireless geophone network architecture based on the IEEE 802.11ad standard has been evaluated under the impact of co-channel interference for a combination of the seismic survey size, number of geophones, data generation rate, and survey terrain in Saudi Arabia and Texas, USA.  \par
On the basis of statistical models for the path loss and LoS probability, and a cross-layer analytical model for the latency and power consumption, an optimization framework is developed to achieve near-optimal power conservation performance through the FA-PSB scheme. A performance evaluation at a data generation rate of 144~Kbps reveals that the power consumption can be reduced by up to 87\% with a maximum latency of around 2~s at the DCC, as compared to classical operation prescribed by the 802.11ad standard. Although previously proposed approaches based on the IEEE 802.11ac standard are feasible for low-rate applications such as earthquake detection and seismic quality control, the proposed use of the IEEE 802.11ad standard is far more suitable for data-intensive real-time seismic acquisition. \par
The power conservation performance of the FA-PSB scheme can be further enhanced by utilizing a higher value for $K_{max}$, which could be tweaked in vendor-specific implementations to support larger A-MPDU frame sizes. Additionally, data compression techniques can be incorporated to reduce the overall packet arrival rate at the WGNs, which in turn would enable the FA-PSB scheme to conserve more power. A reduction in the power consumption translates to a reduced cost in terms of the equipment weight, transportation, and manpower. The proposed architecture also offers a low-cost alternative to current seismic data acquisition systems by eliminating cable and reducing the overall power consumption. Furthermore, the FA-PSB scheme can find application in cellular backhaul and large-scale sensor networks for effective power conservation.
\appendices 
\section{}
\label{appendix:cci}
For a pair of co-channel cells, the probability of the occurrence of CCI is affected by both the LoS probability and the probability of overlap in the data transmission periods. Considering the $l'^{th}$ sub-link, let $m'$ and $n'$ denote the sub-links that are the first-tier co-channel cells. 
\begin{smalleralign}[\normalsize]
p_{l'}^{cci} &= 1-\left(1-p^{LoS}_{l',m'} \cdot p^{ov}_{l',m'}\right) \left(1-p^{LoS}_{l',n'} \cdot p^{ov}_{l',n'} \right)
\end{smalleralign}
where $p^{LoS}_{l',m'}$ and $p^{ov}_{l',m'}$ denote the LoS and overlap probability respectively, between the $l'^{th}$ and $m'^{th}$ sub-links. The value for $p^{LoS}_{l',m'}$ can be determined using the model in Section~\ref{section:statistical}. Following the analysis technique used in~\cite{overlapProb}, an expression for $p^{ov}_{l',m'}$ is obtained in~\eqref{eq:overlapProb}.
\begin{smalleralign}[\normalsize]
p^{ov}_{l',m'} &= \begin{cases}
(t_{m,d}+t_{l,d})/(K_{m}/\lambda_{m}-\delta) ~,&~ t_{m,sl}>t_{l,d} \\
[1 + (t_{m,d}+t_{l,d})/(K_{m}/\lambda_{m}-\delta) ]/2 ~,&~ t_{m,sl} \leq t_{l,d} 
\end{cases} \label{eq:overlapProb}
\end{smalleralign}
\section{}
\label{appendix:relays}
Let $p_{los}(d)$ denote the LoS probability for a link distance $d$. Additionally, let $\mathcal{R}$ denote a random variable that represents the number of relays that are present in the link under consideration. The mean number of relays is then given by
\begin{smalleralign}[\normalsize]
\mathbb{E}(\mathcal{R}) &= \sum\limits_{r=0}^{\infty} r \cdot p_{r} \label{eq:relayprod}
\end{smalleralign}
where $p_{r}=\Pr(\mathcal{R}=r)$. Define $q_{r} \triangleq p_{los}\left(\frac{d}{r+1}\right)$. When $r=0$, it is intuitive that $p_{0}=q_{0}$. When $r=1$, it is implied that there is no LoS for the entire link distance $d$, and that LoS conditions exist for both the sub-links. Hence, $p_{1} = q_{1}^{2}(1-q_{0})$. The following generalisation can be made for $p_{r}$.
\begin{smalleralign}[\normalsize]
p_{r} &= \begin{cases} 
q_{0} &~~ r = 0 \\
q_{r}^{r+1} \prod\limits_{r'=0}^{r-1} (1-q_{r'}) &~~ r \geq 1
\end{cases} 
\end{smalleralign}
Note that \eqref{eq:relayprod} represents a convergent series since $q_{r}=1$ for some finite $r$ i.e., LoS conditions exist for some arbitrarily small value of the sub-link distance. Hence, the minimum required number of relays in \eqref{eq:minRelays} can be expressed as $\lceil \mathbb{E}(\mathcal{R}) \rceil$ , where $\lceil \bullet \rceil$ denotes the ceiling function.
\section{}
\label{appendix:convex}
Taking the second derivative of $\Phi$ with respect to $K_{l}$, where $K_{l}\in\mathbb{K}$,
\begin{smalleralign}[\normalsize]
\dfrac{\partial^{2} \Phi}{\partial K_{l}^{2}} &= \dfrac{\partial^{2}}{\partial K_{l}^{2}} \left\{ \dfrac{1}{\alpha_{1}e^{\beta_{1}K_{l}}-\alpha_{2}e^{-\beta_{2}K_{l}}} \right\} \\
 & \hspace*{-5mm} = \dfrac{e^{K_{l}(\beta_{1}-\beta_{2})}}{\Phi} \left\{ \alpha_{1}^{2}\beta_{1}^{2}e^{K_{l}(\beta_{1}+\beta_{2})} + \alpha_{2}^{2}\beta_{2}^{2}e^{-K_{l}(\beta_{1}+\beta_{2})} + \right. \nonumber  \\ & \left. \alpha_{1}\alpha_{2}(\beta_{1}^{2}+4\beta_{1}\beta_{2}+\beta_{2}^{2}) \right\} \\[5pt]
& \hspace*{-5mm} = \dfrac{e^{K_{l}(\beta_{1}-\beta_{2})}}{\Phi} \left\{ \left( \alpha_{1}\beta_{1}e^{K_{l}(\beta_{1}+\beta_{2})/2} + \alpha_{2}\beta_{2}e^{-K_{l}(\beta_{1}+\beta_{2})/2} \right)^{2} + \right. \nonumber  \\ & \left. \alpha_{1}\alpha_{2} \left( \beta_{1}+\beta_{2} \right)^{2} \right\} \\[5pt]
&\hspace*{-5mm} \geq 0 \hspace*{3cm} \forall ~\alpha_{1},~\beta_{1},~\alpha_{2},~\beta_{2}~\in~\mathbb{R}^{+}
\end{smalleralign}
Since $\partial^{2} \Phi/\partial K_{l}^{2}$ is non-negative $\forall~K_{l}\in\mathbb{K}$, $\Phi$ is convex over $\mathbb{K}$~\cite{Boyd}.
\section{}
\label{appendix:values}
As listed in Table~\ref{tab:rx-sens}, the minimum receiver sensitivity ($\text{Rx}_{min}$) and signal-to-interference-plus-noise ratio ($\text{SINR}_{min}$) values for each of the MCS indices are derived from~\cite{80211-2020,sinr}.
\begin{table}[t!]
\centering
\normalsize
\caption{}
\resizebox{\columnwidth}{!}{
\begin{tabular}{| C{0.14\columnwidth} | C{0.16\columnwidth} | C{0.18\columnwidth} | C{0.14\columnwidth} | C{0.16\columnwidth} | C{0.18\columnwidth} |}
\hline
\textbf{MCS Index} & $\text{Rx}_{min}$ (dBm) & $\text{SINR}_{min}$ (dB) & \textbf{MCS Index} & $\text{Rx}_{min}$ (dBm) & $\text{SINR}_{min}$ (dB) \\ \hline
13 & -66 & 6.95 & 19 & -56 & 17.85 \\ \hline
14 & -64 & 8.15 & 20 & -54 & 19.2 \\ \hline
15 & -63 & 10.05 & 21 & -53 & 20.7 \\ \hline
16 & -62 & 11.25 & 22 & -51 & 22.85 \\ \hline
17 & -60 & 12.65 & 23 & -49 & 24.8 \\ \hline
18 & -58 & 15.8 & 24 & -47 & 26.15 \\ \hline
\end{tabular}
\label{tab:rx-sens}
}
\end{table}
%\newpage
\setstretch{1.05}
\bibliographystyle{IEEEtran}
\bibliography{IEEEabrv,Ref}

%\vskip -20pt plus -1fil

\begin{IEEEbiography}[{\includegraphics[width=1in,height=1.25in,clip,keepaspectratio]{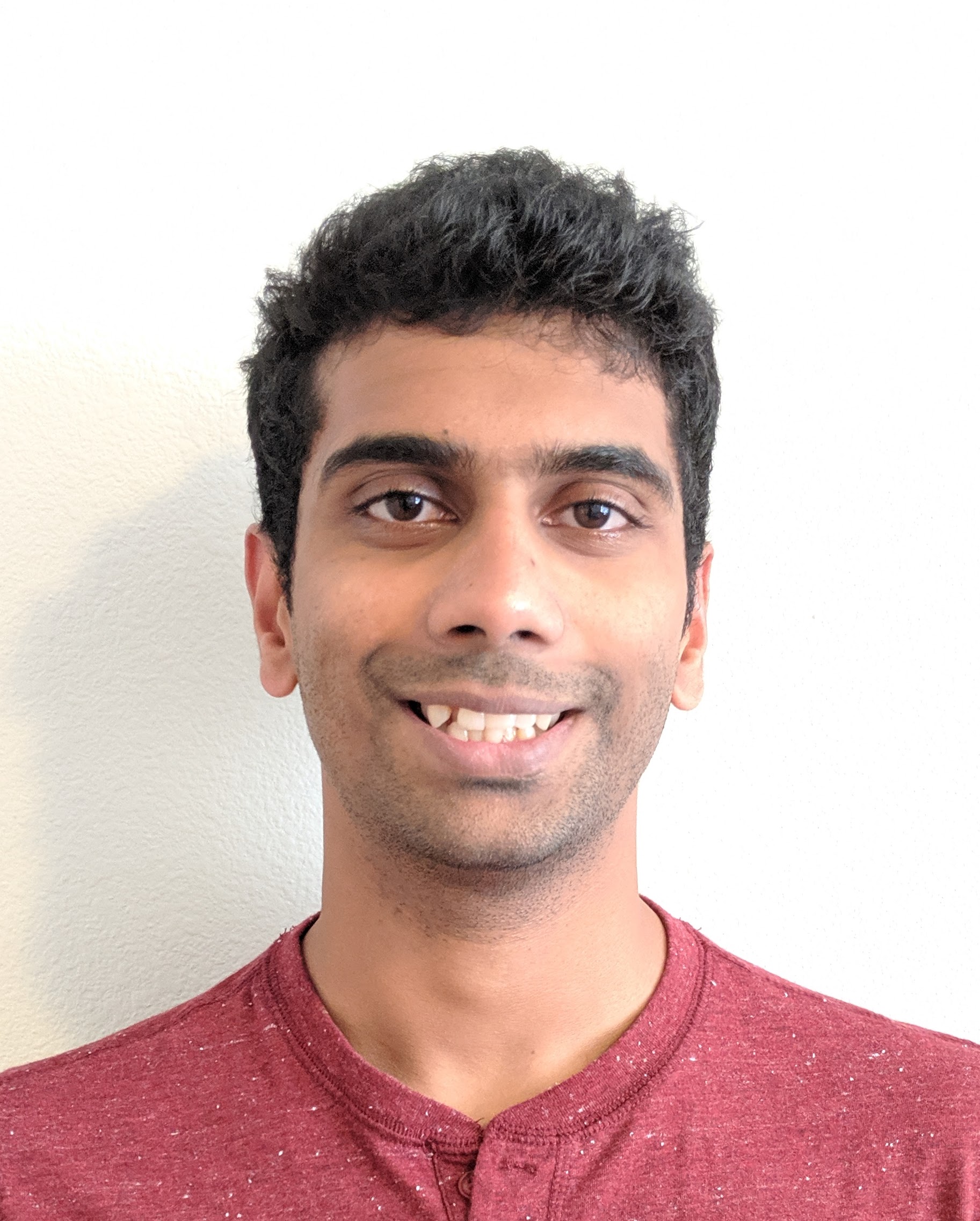}}]{Varun Amar Reddy} received the B.Tech. degree in Electronics and Communication Engineering from the National Institute of Technology Karnataka, Surathkal, India in 2016. He received the Ph.D. degree in Electrical and Computer Engineering from the Georgia Institute of Technology, Atlanta, USA in 2021. His research interests include wireless networking, resource allocation for access protocol design, energy-efficient communications, and next-generation positioning systems. He has been granted one U.S. patent. He has served as a reviewer for IEEE Transactions on Vehicular Technology and various IEEE conferences. 
\end{IEEEbiography}

%\vskip 0pt plus -1fil

\begin{IEEEbiography}[{\includegraphics[width=1in,height=1.25in,clip,keepaspectratio]{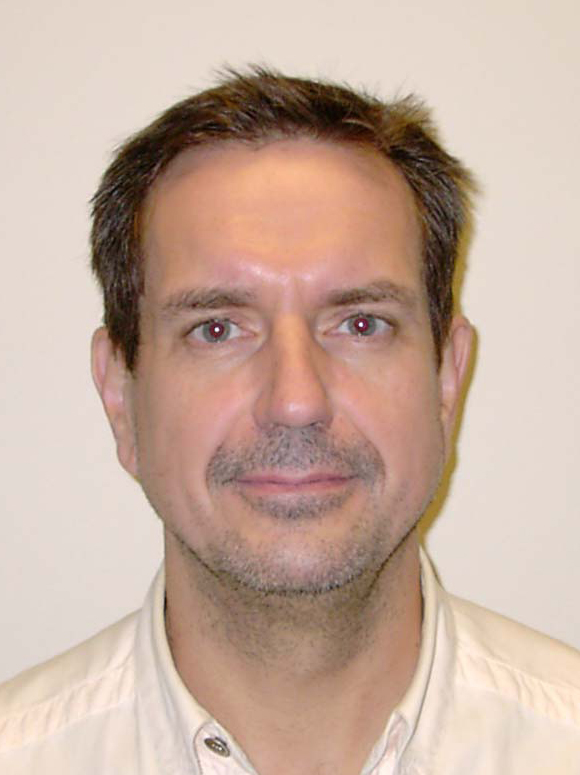}}]{Gordon~L.~St{\"u}ber} (S'81-M'82-SM'96-F'99) received the B.A.Sc. and Ph.D. degrees in electrical engineering from the University of Waterloo, Waterloo, ON, Canada, in 1982 and 1986, respectively. In 1986, he joined the School of Electrical and Computer Engineering, Georgia Institute of Technology, where he is the Joseph M. Pettit Chair Professor in Communications. \par
He is the author of the wireless textbook Principles of Mobile Communication (Kluwer Academic Publishers, 1996, 2/e 2001, 3/e 2011, 4/e 2017). He was a co-recipient of the Jack Neubauer Memorial Award in 1997 for the best systems paper published in the IEEE Transactions on Vehicular Technology and the Neal Shepherd Memorial Best Propagation Paper Award in 2012, for the best propagation paper published in the IEEE Transactions on Vehicular Technology. He became an IEEE Fellow in 1999 ``for contributions to mobile radio and spread spectrum communications.'' He was the recipient of the IEEE Vehicular Technology Society James R. Evans Avant Garde Award in 2003 ``for his contributions to theoretical research in wireless communications,'' the 2007 IEEE Communications Society Wireless Communications Technical Committee Recognition Award ``for outstanding technical contributions in the field and for service to the scientific and engineering communities,'' and the 2017 IEEE ComSoc RCC Technical Recognition Award ``for outstanding research contributions to radio communications,'' and the IEEE Vehicular Technology Society Outstanding Service Award in 2005. He was an IEEE Communication Society Distinguished Lecturer (2007-2008) and IEEE Vehicular Technology Society Distinguished Lecturer (2010-2012). \par
Dr. St{\"u}ber was the Technical Program Chair for the 1996 IEEE Vehicular Technology Conference (VTC'96), the Technical Program Chair for the 1998 IEEE International Conference on Communications (ICC'98), the General Chair of the Fifth IEEE Workshop on Multimedia, Multiaccess and Teletraffic for Wireless Communications (MMT'00), the General Chair of the 2002 IEEE Communication Theory Workshop (CTW'02), General Chair of the Fifth YRP International Symposium on Wireless Personal Multimedia Communications (WPMC'02), and General Co-Chair of the 2019 IEEE 90th Vehicular Technology Conference (VTC2019-Fall). He is a past Editor for Spread Spectrum with the IEEE Transactions on Communications (1993-1998), a past member of the IEEE Communications Society Awards Committee (1999-2002). He was an elected Member-at-Large on the IEEE Communications Society Board of Governors (2007-2009), and is an elected member of the IEEE Vehicular Technology Society Board of Governors (2001-2021).
\end{IEEEbiography}

%\vskip 0pt plus -1fil

\begin{IEEEbiography}[{\includegraphics[width=1in,height=1.25in,clip,keepaspectratio]{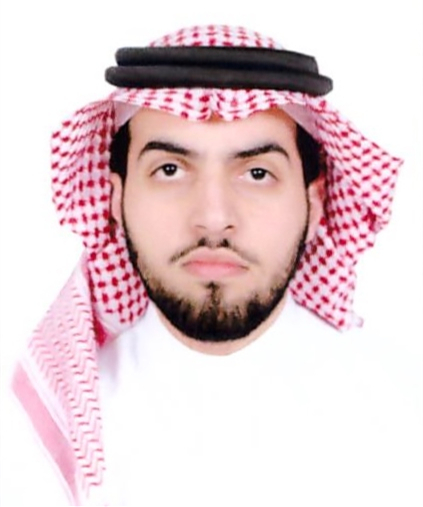}}]{Suhail Al-Dharrab} (GS'07-M'13-SM'20) received the B.Sc. degree in electrical engineering from King Fahd University of Petroleum and Minerals, Dhahran, Saudi Arabia, in 2005, and the M.A.Sc. and Ph.D. degrees in electrical and computer engineering from University of Waterloo, Waterloo, ON, Canada, in 2009 and 2013, respectively. From 2005 to 2007, he was a Graduate Assistant with the Electrical Engineering Department, King Fahd University of Petroleum and Minerals. In 2015, he was a Visiting Professor with the School of Electrical and Computer Engineering, Georgia Institute of Technology, Atlanta, USA. He is currently an Assistant Professor with the Electrical Engineering Department and the Assistant Dean of Research with King Fahd University of Petroleum and Minerals. His research interests span topics in the areas of wireless communication systems, underwater acoustic communication, digital signal processing, and information theory.
\end{IEEEbiography}

%\vskip 0pt plus -1fil

\begin{IEEEbiography}[{\includegraphics[width=1in,height=1.25in,clip,keepaspectratio]{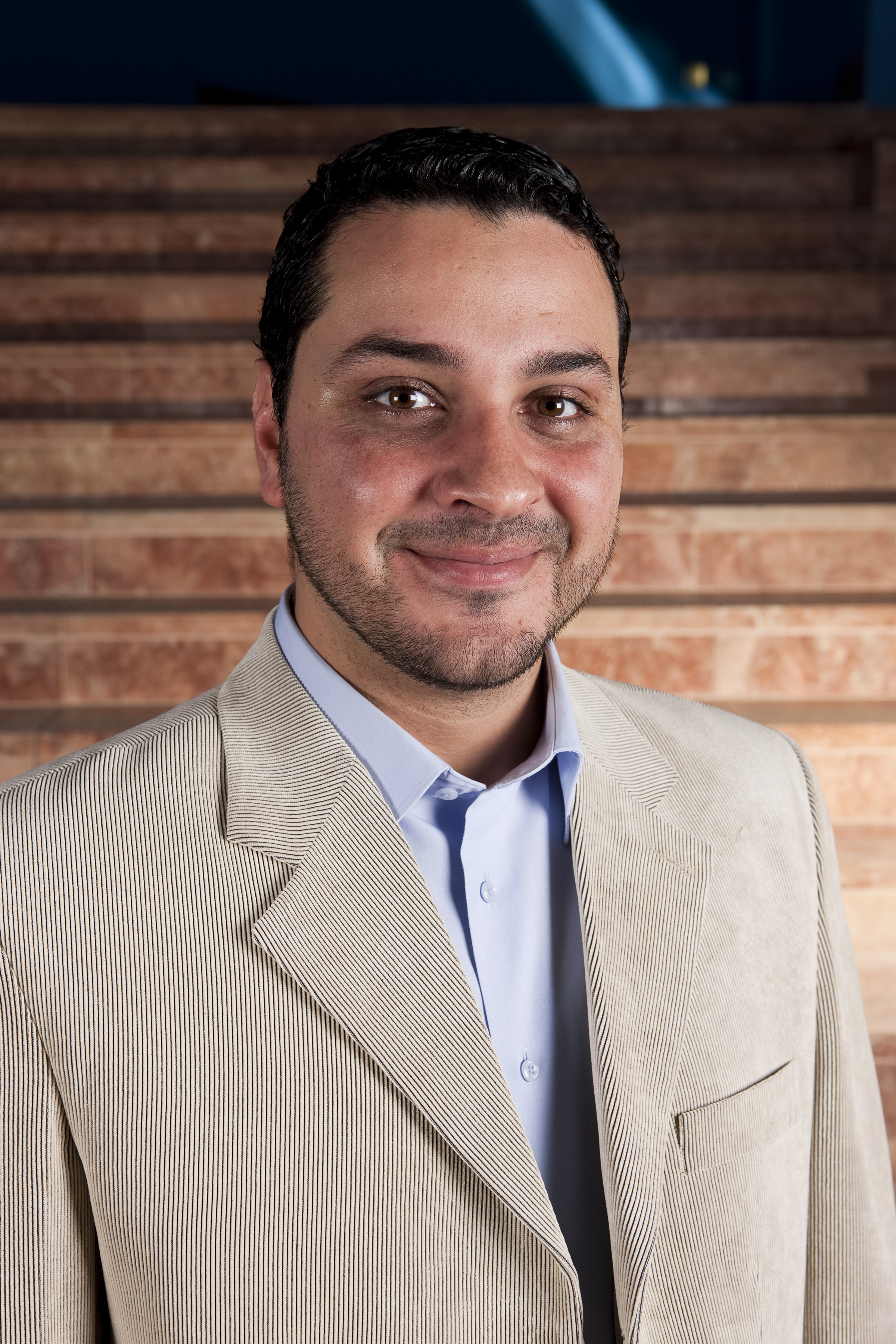}}]{Wessam Mesbah} (M'09-SM'15) received the M.Sc. and B.Sc. degrees (Hons.) in electrical engineering from Alexandria University, Alexandria, Egypt, in 2003 and 2000, respectively, and the Ph.D. degree from McMaster University, Hamilton, ON, Canada, in 2008. From 2009 to 2010, he was a Post-Doctoral Research Fellow with Texas A\&M University, Doha, Qatar. He joined the Electrical Engineering Department, King Fahd University of Petroleum and Minerals, in 2010, where he is currently an Associate Professor. His research interests include cooperative communications and relay channels, layered multimedia transmission, wireless sensor networks, multiuser MIMO/OFDM systems, cognitive radio, optimization, game theory, and smart grids.
\end{IEEEbiography}

%\vskip 0pt plus -1fil

\begin{IEEEbiography}[{\includegraphics[width=1in,height=1.25in,clip,keepaspectratio]{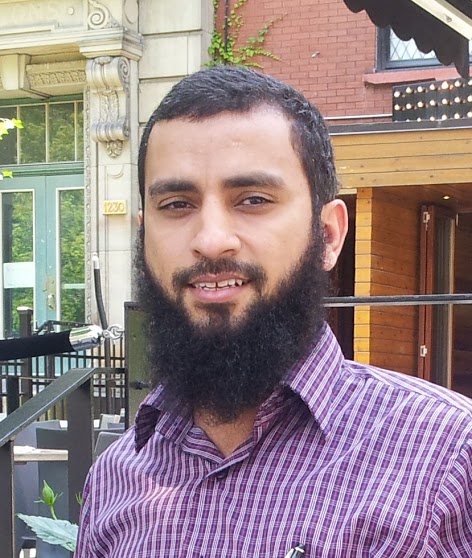}}]{Ali Hussein Muqaibel} (SM) received the B.Sc. and M.Sc. degrees from the King Fahd University of Petroleum and Minerals (KFUPM), Dhahran, Saudi Arabia, in 1996 and 1999, respectively, and the Ph.D. degree from the Virginia Polytechnic Institute and State University, Blacksburg, VA, USA, in 2003. During his study at Virginia Tech, he was with both the Time Domain and RF Measurements Laboratory and the Mobile and Portable Radio Research Group. He was a Visiting Associate Professor with the Center of Advanced Communications, Villanova University, Villanova, PA, USA, in 2013, a Visiting Professor with the Georgia Institute of Technology in 2015, and a Visiting Scholar with the King Abdullah University for Science and Technology (KAUST), Thuwal, Saudi Arabia, in 2018 and 2019. He is currently a professor with the Electrical Engineering Department, KFUPM. He has authored two book chapters and over 130 articles. His research interests include direction of arrival estimation, through-wall-imaging, localization, channel characterization, and ultra-wideband signal processing. He was a recipient of many awards in the excellence in teaching, advising, and instructional technology.
\end{IEEEbiography}

\end{document}